\documentstyle[12pt]{article}
\newlength{\pubnumber} 

\catcode`\@=11
\@addtoreset{equation}{section}
  
\def\section{\@startsection{section}{1}{\z@}{3.5ex plus 1ex minus .2ex}
 {2.3ex plus .2ex}{\large\bf}}
\def\subsection{\@startsection{subsection}{2}{\z@}{2.3ex plus .2ex}
 {2.3ex plus .2ex}{\bf}}

    \renewcommand{\baselinestretch}{1.4}

\def\X{{\bf X}}

\begin{document}

\begin{titlepage}
\samepage{
\setcounter{page}{1}
\rightline{UFIFT-HEP-97-29 ; UMD--PP--98--54}
\rightline{{\tt hep-ph/9712516}, {December 1997}}
\vfill
\begin{center}
 {\Large \bf  A Family--Universal Anomalous $U(1)$ \\in String
     Models as the Origin of\\ Supersymmetry Breaking and Squark Degeneracy\\}
\vspace{.12in}
 {\large Alon E. Faraggi$^1$
	   $\,$and$\,$ Jogesh C. Pati$^{2}$
			\\}
\vspace{.12in}
 {\it  $^{1}$ Department of Physics,
		University of Florida, Gainesville, FL 32611 USA\\}
\vspace{.05in}
 {\it  $^{2}$  Department of Physics,
               University of Maryland, College park, MD 20742\\}
\end{center}
\vspace{.1in}
\begin{abstract}
  {\rm
Recently a promising mechanism for supersymmetry breaking that utilizes
both an anomalous $U(1)$ gauge symmetry and an effective mass term
$m\sim1$TeV of certain relevant fields has been proposed. In this
paper we examine whether such a mechanism can emerge in superstring
derived free fermionic models. We observe that certain three
generation string solutions, though not all, lead to an anomalous
$U(1)$ which couples universally to all three families.
The advantages of this three--family universality of $U(1)_A$,
compared to the two--family case, proposed in earlier works,
in yielding squark degeneracy, while avoiding radiative breaking of
color and charge, are noted. The root cause of the flavor universality
of $U(1)_A$ is the cyclic permutation symmetry that characterizes the
$Z_2\times Z_2$ orbifold compactification with standard embedding,
realized in the free fermionic models by the NAHE set. It is shown
that non--renormalizable terms which contain hidden--sector condensates,
generate the required suppression of the relevant mass term $m$, compared
to the Planck scale. While the $D$--term of the family universal $U(1)_A$
leads to squark degeneracy, those of the family dependent $U(1)$'s,
remarkably enough, are found to vanish for the solutions considered,
owing to minimization of the potential.
}
\end{abstract}
\vfill
\smallskip}
\end{titlepage}

\setcounter{footnote}{0}

\def\beq{\begin{equation}}
\def\eeq{\end{equation}}
\def\beqn{\begin{eqnarray}}
\def\eeqn{\end{eqnarray}}
\def\AEF{A.E. Faraggi}
\def\NPB#1#2#3{{\it Nucl.\ Phys.}\/ {\bf B#1} (19#2) #3}
\def\PLB#1#2#3{{\it Phys.\ Lett.}\/ {\bf B#1} (19#2) #3}
\def\PRD#1#2#3{{\it Phys.\ Rev.}\/ {\bf D#1} (19#2) #3}
\def\PRL#1#2#3{{\it Phys.\ Rev.\ Lett.}\/ {\bf #1} (19#2) #3}
\def\PRT#1#2#3{{\it Phys.\ Rep.}\/ {\bf#1} (19#2) #3}
\def\MODA#1#2#3{{\it Mod.\ Phys.\ Lett.}\/ {\bf A#1} (19#2) #3}
\def\IJMP#1#2#3{{\it Int.\ J.\ Mod.\ Phys.}\/ {\bf A#1} (19#2) #3}
\def\nuvc#1#2#3{{\it Nuovo Cimento}\/ {\bf #1A} (#2) #3}
\def\etal{{\it et al,\/}\ }
\hyphenation{su-per-sym-met-ric non-su-per-sym-met-ric}
\hyphenation{space-time-super-sym-met-ric}
\hyphenation{mod-u-lar mod-u-lar--in-var-i-ant}
\def\l{\langle}
\def\r{\rangle}
\def\vsq#1{\vert\l{#1}\r\vert^2}
\def\p23{\vsq{{\bar\Phi}_{23}}}
\def\v32{\vsq{V_3}}
\def\h18{\vsq{H_{18}}}
\def\anomaly{{{g^2}\over{16\pi^2}}{1\over{2\alpha^\prime}}}

\setcounter{footnote}{0}
\section{Introduction}

Understanding the origin of (i) supersymmetry breaking
and simultaneously (ii) of the extreme degeneracy in the masses
of the squarks in at least the first two families, as inferred
from the miniscule strengths of the $K^0-{\bar K}^0$ transition,
is still among the important unsettled issues in particle physics.
Equally important is understanding the large hierarchy between the
Planck scale and the SUSY breaking mass splitting $\delta m_s$,
reflected by the ratio $\delta m_s/M_{\rm Planck}\sim10^{-15}$.

Several mechanism have been proposed to implement SUSY--breaking.
These include the ideas of: (i) gaugino condensation in the
hidden sector \cite{gauginocond}; (ii) dilaton dominated SUSY
breaking (DDSB) \cite{DDSB}, (iii) gauge mediated SUSY breaking (GMSB)
\cite{GMSB}; whose intrinsic origin is delegated to an unknown
mechanism involving an effective singlet field, which couples
to a set of messenger particles; and (iv) SUSY-- breaking, induced
through joint effects of an anomalous $U(1)$ gauge interaction
and effective mass--terms of certain relevant fields, which
carry the anomalous $U(1)$ charge \cite{fh1,dvali,ua1}.
The mass terms in case (iv) represent the scale of
SUSY--breaking mass splitting $\delta m_s$, and thus, on
phenomenological grounds, they must be of order 1 TeV.

Among these, the mechanism of DDSB
automatically yields squark degeneracy at the tree level.
It however has the problem of possible color and charge breaking
(see {\it  e.g.} the last paper in ref. \cite{DDSB}).
The GMSB yields squark degeneracy, provided that the
superpotential Yukawa interactions of the messenger fields with
the Standard Model fields are suppressed. The existence of
such messenger fields in superstring
derived models was proposed \cite{gmssb}.
That of the anomalous $U(1)$ can yield the desired
degeneracy provided that it couples universally to at least
the first two families, which is assumed in refs. \cite{dvali,mohapatra}.
Short of deriving any of these from an underlying theory,
such as superstring theory, however, the scale of SUSY--breaking
mass splittings $\delta m_s$ as well as the choice
of fields and of their quantum numbers are rather arbitrary.
They must therefore be put in by hand.

It is thus of great interest to examine whether any of
these mechanisms could in fact emerge from within
a superstring theory. Now, phenomenologically viable,
solutions of string theory invariably
do indeed contain an anomalous $U(1)$ as a generic
feature (see {\it e.g.} refs. \cite{revamp,fny,alr,price,eu,top,lykken},
as examples of models based on the free fermionic construction \cite{fff}.
We wish therefore to explore in this note the viability
of supersymmetry breaking through an anomalous $U(1)$
in the context of such string derived solutions.
In particular, we examine whether they can yield
either a two or a three--family universal anomalous $U(1)$ that would lead
to squark mass degeneracy (following SUSY--breaking) on the one hand,
and yet would not conflict with the observed
hierarchy in the masses of the fermions on the other hand;
and whether these solutions can also yield non--vanishing but
strongly suppressed mass terms $m<< M_{\rm Planck}$ of certain relevant
fields, which are essential to trigger SUSY--breaking.
The smallness of $m$ compared to the Planck mass
would then account for the large hierarchy between
$\delta m_s$ and $M_{\rm Planck}$.

Given that string theory yields a vast set of solutions
at the tree level and that no guiding principle
is available to choose between them, it is, of course, still
premature to take too seriously any specific
solution. Yet certain generic features of a class
of solutions, related especially to their
{\it symmetry properties}, may well survive in the
final picture. With this in mind and for
concreteness, we examine the issues
noted above
within a specific class of string--derived solutions, which are obtained
in the free fermionic formulation \cite{fff},
and yield non--GUT standard model--like gauge
symmetries with three generations \cite{eu,top}.
Later, we will comment
on the issue of flavor universality of the anomalous $U(1)$ in some other
solutions such as those of refs. \cite{revamp} and
\cite{alr}. A priori motivations for considering
the class of solutions obtained in refs \cite{eu,top}
are that (a) they seem capable of generating qualitatively
the right--texture for fermion masses and mixings; (b)
they provide a natural doublet--triplet
splitting mechanism because of their non--GUT character
and (c) they also possess extra gauge symmetries, beyond
SUSY GUTs, which, together with the allowed pattern of VEVs,
safeguard proton stability from all
potential dangers \cite{custodial,jp}, including those
which may arise from higher dimensional operators and
from the exchange of color triplets in the heavy tower
of string states. These extra symmetries also turn
out to be helpful in suppressing $\nu_L-{\tilde H}$
mixing operator \cite{nhmix}. Last but not the least,
having their origin in a string theory, they of course
satisfy gauge coupling unification in spite of their
non--GUT character \cite{ginsparg}.
The obvious question is whether this class of string
solutions also permits supersymmetry breaking at the electroweak scale
through an anomalous $U(1)$, while preserving family--universality
in squark masses.

In section 2, we observe that the desired family
universality of the anomalous $U(1)$ is by no means
a general property of string solutions, but it holds
in the class of solutions obtained
in ref. \cite{eu} and \cite{top}.
We point out the root cause why it holds for this class.
In section 3 and 4, we discuss supersymmetry breaking
and generation of relevant mass--terms for these solutions.
We show in section 4 that there exist solutions in this class
which yield operators of dimension $n\ge4$, that induce
highly suppressed relevant mass terms, $\sim(1/2-50){\rm TeV}$.
These mass terms, together
with the anomalous $U(1)$, induce SUSY breaking.
Thus these solutions have the potential for explaining
(a) supersymmetry breaking; (b) gauge hierarchy and (c)
squark degeneracy.

An issue of special concern is that
string solutions invariably possess a
host of other $U(1)$s, which are family dependent,
and contributions from their $D$--terms, if non--vanishing,
could potentially spoil squark degeneracy.
We show in section 3 that there exist solutions
for which, the contributions from the
undesirable $D$--terms, remarkably enough,
vanish owing to a minimization of the potential.
In short, the class of string solutions
considered here, though by no means unique,
possesses three non--trivial and highly desirable
features: (i) a family universal anomalous $U(1)$,
(ii) suitably suppressed mass terms of relevant
fields which trigger SUSY breaking, and (iii)
vanishing of family dependent
$D$--term contributions. In section 5, we mention certain
features of phenomenological interest. In particular
string solutions of ref. \cite{eu} or \cite{top} lead to
approximate squark degeneracy for {\it all three families}.
This is in contrast to the case of ref. \cite{dvali} and \cite{mohapatra},
where the degeneracy holds (because of the choice of
the anomalous charge) only for the first two families.
Advantages of three compared to
two family squark--degeneracy
in avoiding radiative color and electric charge breaking --
is noted. We also point out that the solution
of ref. \cite{top} leads to intra--family
sfermion degeneracy ({\it i.e.} $m_{\tilde q_L}=m_{\tilde u_R}=
m_{\tilde d_L}=m_{\tilde L}$, etc.), whereas that of ref. \cite{eu}
leads to considerable splitting between the members of a family.
In the concluding section, we make some general remarks about the prospect
of supersymmetry--breaking through $U(1)_A$, and the possibility of
other contributions to supersymmetry breaking.

\setcounter{footnote}{0}
\section{A family universal anomalous $U(1)$ in a class of string solutions}

We begin by recalling certain salient features of the solutions
based on the free fermionic formulation \cite{fff}. They are
defined by a set of boundary condition basis vectors
and the associated one--loop GSO projection
coefficients. The massless states are obtained by
applying the generalized GSO projections. Each massless
state defines a vertex operator and the cubic and
higher order terms in the superpotential are obtained
by calculating the correlators between the vertex
operators \cite{cvetic,kln}.

The specific class of solutions \cite{eu,top} which
we examine here are generated by a set of eight
boundary condition basis vectors. The first five
of these, denoted by $\{{\bf1}, S, b_1,b_2,b_3\}$,
constitute the so--called NAHE set \cite{revamp,fn}.
They are common to a large class of viable string
solutions, including those of refs. \cite{revamp}
\cite{alr}, and \cite{price}. The properties of the NAHE
set are crucial to understanding how the required
flavor universality of the anomalous $U(1)$ may arise
in certain free fermionic solutions. We therefore
refer the reader to refs. \cite{revamp,fn} for
definition of these basis vectors and their
detailed properties. Here we note only certain salient
features. The vectors $\{{\bf1}, S\}$ give rise
to a solution with $N=4$ space--time supersymmetry
and $SO(44)$ gauge symmetry. The vectors $b_1$, $b_2$
and $b_3$ break $N=4$ to $N=1$ and $SO(44)$ to
$SO(10)\times E_8\times SO(6)^3$, where $SO(10)$ is
identified with the GUT symmetry containing
the Standard Model. Each of the vectors $b_1$,
$b_2$ and $b_3$ produces 16 multiplets, each of
which is a 16 of $SO(10)$; thus there are altogether
48 generations. The sixteen generations produced
by each $b_j$ are charged with respect to only
one of the $SO(6)_j$ -- symmetries,
which is why the $SO(6)_j$ -- symmetries provide the
origin of flavor symmetries. Note that at this stage,
there is a complete {\it permutation symmetry} between the
sectors $b_1$, $b_2$ and $b_3$,
which is reflected in the full set of gauge interactions
as well as in the superpotential.
It is this permutation symmetry which leads
to family universality of the anomalous $U(1)$
in some models.

It is important to note that the NAHE set corresponds to
a $Z_2\times Z_2$ orbifold compactification. This
seemingly apparent observation has far reaching
phenomenological implications.
The focus in this paper is on SUSY breaking and
squark degeneracy. The correspondence of the
NAHE set with $Z_2\times Z_2$ orbifold compactification is
best illustrated by adding to the NAHE set the boundary
condition basis vector $\X$ with periodic boundary conditions
for the world--sheet
fermions $\{{\bar\psi}^{1,\cdots,5},{\bar\eta}^{1,2,3}\}$,
and antiperiodic boundary conditions for all others. With a
suitable choice of the generalized GSO projection coefficients
the $SO(10)$ gauge group is enhanced to $E_6$. The $SO(6)^3$
symmetries are broken to $SO(4)^3\times U(1)^3$. One combination
of the $U(1)$ symmetries is embedded in $E_6$. The gauge group in this case
would be $E_6\times E_8\times SO(4)^3\times U(1)^2$.
This extended NAHE set then corresponds to
a $Z_2\times Z_2$ orbifold with the standard embedding
of the gauge connection \cite{foc}.
The three sectors
generated by $b_1$, $b_2$ and $b_3$ are
the three twisted sectors of the orbifold models.
The cyclic permutation symmetry associated with the NAHE
set is thus simply the symmetry between the three
twisted sectors of the $Z_2\times Z_2$ orbifold, with standard embedding.
The permutation symmetry also applies to the spectrum
that arises from the untwisted sector, including the moduli.
The phenomenological motivation for this symmetry
will become apparent in the context of supersymmetry breaking and squark
degeneracy.
Whether this symmetry is unique to the $Z_2\times Z_2$
orbifold compactification is an open question.

The next stage in the construction of viable
solutions is the introduction of additional
boundary condition basis vectors, which reduce
the number of chiral generations from forty-eight
to three, barring possible vector--like multiplets.
These also break $SO(10)$ to one of its subgroups --
{\it e.g.} $SU(5)\times U(1)$ \cite{revamp},
$SO(6)\times SO(4)$ \cite{alr} or $SU(3)\times SU(2)\times U(1)^2$
\cite{fny,eu,top}. The hidden $E_8$ symmetry is typically
also broken to one of its subgroups, and the horizontal
$SO(6)^3$ symmetry breaks typically to Abelian factors
of $U(1)^n$, where $n(\ge3)$ varies between the solutions.
As we discuss below, the permutation symmetry of the
full set of gauge interactions with respect to the
three chiral families is retained for solutions
of the type presented in refs \cite{eu} and \cite{top},
in spite of the introduction of three
additional boundary condition basis vectors
(beyond the NAHE set). But this is not the
case for the solutions of refs. \cite{revamp}
and \cite{alr}. The reason for this difference
is that in the case of ref. \cite{eu} and \cite{top}
the three chiral families have their origin entirely
in the sectors $b_1$, $b_2$ and $b_3$ respectively,
and there are no additional vector--like families. By contrast,
for the cases of refs. \cite{revamp} and \cite{alr},
owing to the nature of the additional boundary condition
basis vectors, there are vector--like multiplets in
addition to the three chiral families, and the latter
do not all arise from the sectors $b_1$, $b_2$ and $b_3$
respectively.

In summary, the NAHE set naturally gives
rise to the permutation symmetry of the three families,
or rather three groups of families, both in the gauge as
well as in the superpotential sector. This symmetry
need not, however, be retained in general in the
presence of additional boundary condition basis vectors,
beyond the NAHE set. It is intriguing that the stated
symmetry is fully retained in the gauge sector (though
it is partially lost in the superpotential, see below)
for solutions of the type presented in \cite{eu} and \cite{top}.
As we will show, this permutation symmetry in the gauge sector
guarantees family--universality of the anomalous $U(1)$.

As concrete examples, we consider the solutions of both
ref. \cite{eu} and ref. \cite{top}, which we will refer
to as solutions I and II respectively.
They are very similar for most purposes, yet they
possess certain crucial differences.
The gauge symmetry in both cases, arising from the NS sector,
after application of all the GSO projections,
has the following form, at the string scale:
\beq
{\cal G}=\left[SU(3)_C\times SU(2)_L\times U(1)_{B-L}
\times U(1)_{T_{3_R}}\right]\times\left[G_M={\prod_{i=1}^6}
U(1)_i\right]\times G_H
\label{gg}
\eeq
Here, $U(1)_i$ denote six horizontal flavor symmetries,
which descend from $SO(6)^3$, and act non--trivially on the
three chiral families, Higgs multiplets as well as Hidden
matter states. In both cases, $G_H=SU(5)_H\times SU(3)_H\times U(1)_H^2$,
is the gauge symmetry of the hidden sector.
In the model of Ref. \cite{top} additional space--time vector
bosons arise from the sector ${\bf1}+\alpha+2\gamma$
and enhance the $SU(3)_C$ gauge group to $SU(4)_C$ \cite{masip}
(which we use in tables 4,5 and 6).
This enhancement, however, does not affect our discussion here.

The massless spectrum of solution I (ref. \cite{eu}), together
with the quantum numbers of the respective states,
is exhibited in tables 1,2 and 3. The spectrum
includes the three generations that arise
from the sectors $b_1$, $b_2$ and $b_3$, the Higgs
like multiplets $h_{1,2,3}$, $h_{45}$ and their
conjugates, the color triplets $(D_{45},{\bar D}_{45})$,
the $SO(10)$ singlets $\Phi_{1,2,3}^\pm$, $\Phi_{45}$,
$\Phi_{12}$, $\Phi_{13}$, $\Phi_{23}$ and their conjugates,
and the hidden sector multiplets $(V_i, {\bar V}_i, T_i
{\bar T}_i)_{i=1,2,3}$. We see that each generation that
arises from the sector $b_j$, is charged with respect
to two of the $U(1)$'s -- {\it i.e.} $U(1)_{R_j}$ and
$U(1)_{R_{j+3}}$. For each right--moving gauged
$U(1)$ symmetry, there is a corresponding left--moving
global $U(1)$ symmetry, denoted by $U(1)_{L_j}$ and
$U(1)_{L_{j+3}}$, and the states from each sector
$b_j$ are charged with respect to two of these
global symmetries.

The spectrum for solution II (Ref. \cite{top})
is very similar. The main difference is that the six
$SO(10)$--singlets $\Phi^\pm_{1,2,3}$ are replaced
by only two fields $\Phi_{1,2}$, while $h_{45}$
is accompanied by an additional doublet $h_{45}^\prime$
and $\Phi_{45}$ by $\Phi_{45}^\prime$, and the color triplets
$(D_{45}, {\bar D}_{45})$ are absent.
These are listed in tables 4, 5 and 6.

While the subgroup $SU(3)_C\times SU(2)_L\times
U(1)_{B-L}\times U(1)_{T_{3_R}}$ of $SO(10)$ treats
all three families universally, it is easy to see from
table 1 that the pairs $(U_1,U_4)$, $(U_2,U_5)$
and $(U_3,U_6)$, respectively couple to families 1,2 and 3
in an {\it identical fashion}. Thus, on the one hand, these
six $U(1)$ symmetries, having their origin in $SO(44)$,
distinguish between the three families, unlike a GUT
symmetry like $SO(10)$; thereby they serve as the origin
of flavor symmetries, which are needed to explain
the hierarchical Yukawa couplings of the three families
(see below). On the other hand, as stated before, they
preserve the full permutation symmetry with respect
to the three families.

It is easy to check that solution I (ref. \cite{eu}) contains six
anomalous $U(1)$ symmetries: ${\rm Tr} U_1=
{\rm Tr} U_2={\rm Tr} U_3=24,{\rm Tr} U_4= {\rm Tr} U_5=
{\rm Tr} U_6=-12$. These can be expressed by one anomalous
combination which is unique
and five non--anomalous ones\footnote{
The normalization of the different $U(1)$ combinations is fixed
by the requirement that the conformal dimension of the
massless states still gives ${\bar h}=1$ in the new basis.
We remark in advance that the proper normalization
must be taken as it affects the minimization of the potential (see below).}:
\beq
U_A={1\over{\sqrt{15}}}(2 (U_1+U_2+U_3) - (U_4+U_5+U_6))~;~ {\rm Tr} Q_A=
{1\over{\sqrt{15}}}180~.
\label{u1a}
\eeq
One choice for the five anomaly--free combinations is
given by
\beqn
{U}_{12}&=& {1\over\sqrt{2}}(U_1-U_2){\hskip .5cm},{\hskip .5cm}
{U}_{\psi}={1\over\sqrt{6}}(U_1+U_2-2U_3),\label{u12upsi}\\
{U}_{45}&=&{1\over\sqrt{2}}(U_4-U_5){\hskip .5cm},{\hskip .5cm}
{U}_\zeta ={1\over\sqrt{6}}(U_4+U_5-2U_6),\label{u45uzeta}\\
{U}_\chi &=& {1\over{\sqrt{15}}}(U_1+U_2+U_3+2U_4+2U_5+2U_6).
\label{uchi}
\eeqn
Note that the anomalous $U(1)$,
containing the {\it sums} of $U_{1,2,3}$
and $U_{4,5,6}$
is universal with respect to all
three families. {\it This flavor universality
of the anomalous $U(1)$
is thus a consequence of the family permutation symmetry of
the six $U(1)$--interactions, mentioned above.}
Of the anomaly free combinations $U_{12}$, $U_\psi$,
$U_{45}$, and $U_\zeta$ are clearly family
dependent, but $U_\chi$ is family universal.

It is worth noting that while solution II
(ref. \cite{top}) differs in detail from solution I
as regards its spectrum of Higgs multiplets and
the $SO(10)$ singlets, its gauge interactions  nevertheless possess
the full permutation symmetry with respect to the three families
just like solution I. In this case, however, there are only three
anomalous symmetries $U_{1,2,3}$, which can be expressed
by one anomalous and two anomaly free combinations :
\beqn
U_A &=& {1\over\sqrt{3}}(U_1+U_2+U_3) ~;~~~~~~~~~
{\rm Tr Q_A}={1\over\sqrt{3}}72 \label{sol2ua}\\
U_{12} &=& {1\over\sqrt{2}}(U_1-U_2) ~~~~;~~~~
U_\psi={1\over\sqrt{6}}(U_1+U_2-2U_3)
\label{sol2u1com}
\eeqn
Note that the anomalous $U(1)$ is again family--universal,
though $U_{12}$ and $U_\psi$ are not.

We next examine the superpotential and the issue of generating
relevant mass terms which would trigger SUSY--breaking in the
string solution of ref. \cite{eu}.

\setcounter{footnote}{0}
\section{Superpotential and SUSY breaking}

The relevant terms in the cubic level superpotential
of solutions I and II are given by
\beqn
W&=&[
{u_{L_1}^c}Q_1{\bar h}_1+{N_{L_1}^c}L_1{\bar h}_1+
{u_{L_2}^c}Q_2{\bar h}_2+{N_{L_2}^c}L_2{\bar h}_2+
{u_{L_3}^c}Q_3{\bar h}_3+{N_{L_3}^c}L_3{\bar h}_3]\nonumber\\
&+&
    [{{h_1}{\bar h}_2{\bar\Phi}_{12}}
+    {h_1}{\bar h}_3{\bar\Phi}_{13}
+    {h_2}{\bar h}_3{\bar\Phi}_{23}
+ {\bar h}_1{h_2}{\Phi_{12}}
+ {\bar h}_1{h_3}{\Phi_{13}}
+ {\bar h}_2{h_3}{\Phi_{23}}]+\nonumber\\
&+&
        h_3{\bar h}_{45}\Phi_{45}+
  {\bar h}_3h_{45}{\bar\Phi}_{45}+
  (\Phi_{23}{\bar\Phi}_{13}{\Phi}_{12}
+{\bar\Phi}_{23}{\Phi}_{13}{\bar\Phi}_{12})+ ...
\label{superpot}
\eeqn
Here a common normalization constant ${\sqrt 2}g$
is not exhibited. Note that the Yukawa couplings
given in the first square bracket and effectively
the second bracket as well respect the family permutation
symmetry, which simultaneously permutes the three families
and the Higgs--multiplets
$({\bar h}_1,{\bar h}_2,{\bar h}_3)$,
and likewise the $\Phi_{ij}$'s, but the rest of the
superpotential (including higher order terms), which
is not shown, does not.

Note that owing to the constraints of the flavor
$U(1)_i$--symmetries, which distinguish between the
families, and the Higgs multiplets,
$\bar h_1$ couples at the cubic level of the superpotential
only to family 1 and not to families 2 and 3;
Similarly ${\bar h}_2$ and ${\bar h}_3$ couple only
to families 2 and 3 respectively. Now, for the case of
solution I (ref. \cite{eu}), where contributions of higher
dimensional operators to the Higgs mass matrix have
been analyzed in detail \cite{nrt}, it has been shown
that the pair $h_3$ and $\bar h_3$
necessarily become superheavy since their masses receive
contributions from the cubic level superpotential terms;
and only one pair of doublets -- {\it i.e.} either
(${\bar h}_1,h_{45})$ or $({\bar h}_2,h_{45})$ --
remains light, while the remaining pairs become
medium heavy $(\sim10^{12}{\rm GeV})$.
It is easy to verify that for solution II
(ref. \cite{top}) as well,  $h_3$ and $\bar h_3$
become superheavy \cite{cfm}. The mass
pattern of the remaining Higgs doublets depend on the
structure of the higher dimensional operators
and the allowed
pattern of VEVs. Considering the similarity of the massless
spectrum in the observable sector for the two
cases, however, it seems rather plausible that the Higgs
spectrum for the two cases would be quite similar.
Following Ref. \cite{top,cfm} we will proceed
by assuming that only one pair of Higgs doublets, like
$({\bar h}_1, h_{45})$ or $({\bar h}_2,h_{45})$,
remain light for both solutions I and II,
and that the remaining pairs become medium or superheavy.

Since only the light Higgs scalars acquire VEVs (radiatively),
it would follow, for the Higgs spectrum mentioned above,
{\it that the up--quark member of only one
family {\it i.e.} the top and $\nu_\tau$ would get
masses at the level of cubic terms in $W$}.
The masses of the other quarks and their mixings would arise
through successively higher dimensional operators,
which permit their couplings to the light Higgs--pair.
Thus, a hierarchy in fermion masses and mixings
arises in spite of the permutation symmetry of the cubic
Yukawa couplings
(see ref. \cite{eu} and especially ref. \cite{top}
for details of this discussion).
Thus, ultimately, such a hierarchy has its origin in
two features : (a) the family dependent
$U(1)_i$--symmetries, which force the three families
to have Yukawa couplings with three distinct
Higgs--multiplets, and (b) the spontaneously
generated asymmetric Higgs mass--matrix.

We now turn to the pattern of symmetry breaking below
the string--scale. The anomalous $U(1)_A$ is broken by the
Dine--Seiberg--Witten mechanism \cite{dsw}
in which a potentially large Fayet--Iliopoulos $D$--term
$\xi$ is generated by the VEV of the dilaton field.
Such a $D$--term would, in general, break supersymmetry, unless
there is a direction in the scalar potential
$\hat\phi=\sum\alpha_i\phi_i$ which is $F$--flat and also
$D$--flat with respect to all the non--anomalous gauge symmetries
and in which $\sum Q_A^i\vert\alpha_i\vert^2<0$. If such a direction
exists, it will acquire a VEV, canceling the Fayet--Iliopoulos
$\xi$--term, restoring supersymmetry and stabilizing the vacuum.
The exception to this picture arises if there exist
mass terms $(m)$ for certain relevant fields
carrying anomalous charge; in this case the anomalous
$D$--term and the $F$--terms would necessarily acquire
nonvanishing VEVs that are proportional to $m$ and
SUSY would be broken.

The set of $D$ and $F$ flat constraints, in the absence
of such mass terms, is given by,
\beqn
&& \langle D_A\rangle=\langle D_\alpha\rangle=
\langle D_\beta\rangle=
\langle F_i\equiv
{{\partial W}\over{\partial\eta_i}}\rangle=0\label{dterms}\\
\nonumber\\
&& D_A=\left[K_A+
\sum Q_A^k\vert\chi_k\vert^2+\xi\right]\label{da}\\
&& D_\alpha=\left[K_\alpha+
\sum Q_\alpha^k\vert\chi_k\vert^2\right]~,~\alpha\ne A\label{dalpha}\\
&& \xi={{g^2({\rm Tr} Q_A)}\over{192\pi^2}}M_{\rm Pl}^2
\label{dxi}
\eeqn
Here $\chi_k$ are the fields which acquire VEVs of order
$\sqrt\xi$, while the $K$--terms contain fields $\eta_i$
like squarks, sleptons and Higgs bosons whose
VEVs vanish,
at this scale. $Q_A^k$ and $Q_A^i$ denote the anomalous
and non--anomalous charges, which are listed
in eqs. (\ref{u1a})--(\ref{sol2u1com}) for solutions I and II,
and $M_{\rm Pl}\approx2\times 10^{18}$ GeV denotes the
reduced Planck mass. The solution ({\it i.e.} the choice of fields
with non--vanishing VEVs) to the set of
equations (\ref{dterms})--(\ref{dalpha}),
though nontrivial, is not unique. A few alternative solutions
have been considered in refs. \cite{eu,top,nrt,nhmix}.

As a general guide, note that $\xi$ is positive
and is of order $10^{-2}M_{\rm Pl}^2$. To cancel
the $\xi$--term in $\langle D_A\rangle$,
in the absence of mass terms,
at least one field with negative $Q_A$
must acquire a VEV. A large set of solutions including those
of refs. \cite{eu} and \cite{top} assigns nonzero VEV to
$\Phi_{45}$, which is the field with the largest negative $Q_A$.
If $\Phi_{45}$ (or a suitable alternative), acquiring a VEV,
is charged with respect to one of the other symmetries,
some additional fields must also acquire VEVs, so that
the full set of $\langle D_{A,\alpha,j}\rangle$ must vanish.

To demonstrate how SUSY breaking could arise it is instructive
to consider first a simple pattern of VEVs satisfying eqs.
(\ref{dterms})--(\ref{dalpha}),
for the case of solution II (ref. \cite{top}).
We will subsequently study a more complicated
pattern of VEVs for solution I (ref. \cite{eu}).

$\underline{\hbox{{\bf SUSY breaking in solution II}:}}$

As an instructive example we consider a pattern which
assigns nonzero VEVs of order $\sqrt{\xi}$ to only two fields
\beq
\langle \{\Phi_{45},{\Phi_{45}'}\}\rangle\ne 0
\label{phi4545p}
\eeq
All other fields have zero VEV. The charges $(Q_A,Q_\psi,Q_{12})$
for $\Phi_{45}$ are $(-2,+1,0)$, and those for $\Phi_{45}^\prime$
are $(0,-3,0)$ (see eqs. (\ref{sol2u1com}) and table II).
Their contributions to the respective $D-$terms are
thus given by,
\beqn
D_A &=& {1\over\sqrt{3}}[K_A-2(\vert{\Phi_{45}}\vert^2-
		\vert{\bar\Phi}_{45}\vert^2)+{\hat\xi}]\label{daka}\\
D_\psi &=& {1\over\sqrt{6}}[K_\psi+
		(\vert\Phi_{45}\vert^2-\vert{\bar\Phi}_{45}\vert^2)-3
		\vert{\Phi_{45}}^\prime\vert^2]\label{dpsikpsi}\\
D_{12} &=& {1\over\sqrt{2}}K_{12}
\label{d274}
\eeqn
where $\hat\xi=\sqrt{3}\xi$.
The contribution of $\Phi_{45}$ and $\Phi_{45}^\prime$ to all
other $D$--terms are zero.
The $K$--terms contain fields like squarks, sleptons and
Higgs bosons which have zero VEVs. Although ${\bar\Phi}_{45}$
is assigned zero VEV, its contribution
is still exhibited to demonstrate that it would be forced
to have zero VEV in the presence of a mass term. All the
F and D flat conditions are satisfied at the cubic level of the
superpotential by assigning
\beq
\vert\langle\Phi_{45}\rangle_0\vert^2=3\vert\langle\Phi_{45}^\prime
\rangle_0\vert^2={\hat\xi}/2
\label{sol2vevs}
\eeq
All other VEVs are zero. The subscript zero signifies
that the VEVs are obtained in the zero mass limit.
Now introduce an effective mass--term $m$ $(<<M_{\rm string})$
for $\Phi_{45}$ and ${\bar\Phi}_{45}$
in the superpotential :
\beq
W \supset m\Phi_{45}{\bar\Phi}_{45}
\label{massterm}
\eeq

We will discuss how such a mass term is likely
to arise through higher dimensional operators
in string theory. The effective potential then takes
the form:
\beq
V={g^2\over2}(D_A^2+D_\psi^2+D_{12}^2)+m^2(\vert\Phi_{45}\vert^2
+\vert{\bar\Phi}_{45}\vert^2)
\eeq
For simplicity of writing, we have put just one gauge
coupling; in practice the various gauge couplings would
differ due to the running even if they are equal at the string scale.
It is now easy to verify that minimization of the potential
would lead to a shift in the VEVs of $\Phi_{45}$ and
$\Phi_{45}^\prime$.

The extremum conditions lead to the following constraints:

\beqn
&&
{{\partial V}\over{\partial\Phi_{45}}} = 0\Rightarrow
{(\Phi_{45})^\dagger}
[-2D_A+{D_\psi\over\sqrt{2}}+\sqrt{3}{m^2\over g^2}]=0\label{sol2vevp45}\\
&& {{\partial V}\over{\partial\Phi_{45}^\prime}} = 0 \Rightarrow
{(\Phi_{45}^\prime)^\dagger} [D_\psi]=0\label{sol2vevp45p}\\
&& {{\partial V}\over{\partial{\bar\Phi_{45}}}} = 0\Rightarrow
{({\bar\Phi_{45}})^\dagger}
[2D_A-{D_\psi\over\sqrt{2}}+
\sqrt{3}{m^2\over g^2}]=0
\label{sol2vevp45b}
\eeqn
Here the fields and the $D$--terms to the right of the arrows
stand for the VEVs of the respective
entities. Since $\langle\Phi_{45}\rangle\ne0$, eq. (\ref{sol2vevp45})
clearly shows that $\langle D_A\rangle$ and/or
$\langle D_\psi\rangle$ must be of order
$m^2/g^2$ and thus SUSY is broken. Since
$\langle{\Phi_{45}^\prime}\rangle\ne0$, eq. (\ref{sol2vevp45p}) implies
$\langle D_\psi\rangle=0$. Eq. (\ref{sol2vevp45})
then yields $\langle D_A\rangle=(\sqrt{3}m^2)/(2g^2).$
Substituting this in Eq. (\ref{sol2vevp45b})
we see that $\langle{\bar\Phi_{45}}\rangle$
must remain zero, even with $m\ne0$.
Now, using the expressions for $D_A$ and $D_\psi$
given in Eqs. (\ref{daka}) and (\ref{dpsikpsi}),
we can determine the VEVs
of $\Phi_{45}$ and $\Phi_{45}^\prime$. Thus, we see that,
for the special choice of VEVs given by
Eq. (\ref{sol2vevs}), which provides a solution to the $F$ and $D$--flat
conditions (eqs. (\ref{dterms}--\ref{dxi}))
in the massless limit, the extremum
condition lead to a unique solution for the pattern
of VEVs in the case of finite mass $(m(\Phi_{45})=m(\bar\Phi_{45})\ne0)$ :
\beqn
&& \vert\langle{\Phi_{45}}\rangle\vert^2=
		{{\hat\xi}\over2}-{3\over4}{m^2\over{g^2}}~~~~;~~~~\cr
&& \vert\langle{\Phi_{45}^\prime}\rangle\vert^2={{\hat\xi}\over6}-
				{1\over{4}}{m^2\over{g^2}}~~;~~\cr
&& \vert\langle{\bar\Phi_{45}}\rangle\vert^2=0
\eeqn
\beqn
&& \langle D_A\rangle={\sqrt{3}\over2}{m^2\over {g^2}}~~~~,~~~~\cr
&& \langle D_\psi\rangle=\langle D_{12}\rangle=0
\eeqn
\beqn
&& F(\Phi_{45})=F(\Phi_{45}^\prime)=0~~;~~\cr
&& F({\bar\Phi}_{45})=m\sqrt{{{\hat\xi}\over2}-{3\over4}{m^2\over{g^2}}}
\label{fphi45}
\eeqn
It may be verified that this VEV--pattern in fact minimizes the potential.
Note that for this simple example, the $D$--term of only
the anomalous charge $Q_A$, which is family universal, is
non--zero, {\it but those of the non--universal
charges $Q_\psi$ and $Q_{12}$ vanish
owing to minimization}. This special feature arises because there
are only two fields ($\Phi_{45}$ and $\Phi^\prime_{45}$),
having non--zero VEVs, and they contribute only to two $D$--terms
($D_A$ and $D_\psi$), but not to $D_{12}$. Thus, to start with,
$D_{12}=0$. Furthermore, $\Phi_{45}^\prime$ contributes only to
$D_\psi$, but not to $D_A$. Thus, extremization of V with respect
to $\Phi_{45}^\prime$ forces $D_\psi=0$ as well (see eq. (\ref{sol2vevp45p})).

Vanishing of the
non--universal $D$--terms has the desirable
consequence that squarks of all three families
receive the same contribution to their
masses from the $D$--terms, in spite of the presence of the
non--universal flavor symmetries:
\beq
[m^2_{{\tilde q}_i}]_{D_A}=g^2Q_A^i\langle D_A\rangle=
			Q_A^i({\sqrt{3}\over2}m^2)={m^2\over4}
\label{msqda}
\eeq
Likewise for the sleptons. Here $Q_A^i$ denotes the
anomalous charge of the respective field. For solution II,
$Q_A^i=(Q_1+Q_2+Q_3)^i$ is not only
family--universal, but it is also positive and
the same $(=1/\sqrt{12})$ for all members of a
family. Thus, $\tilde Q_L$, $\tilde d_R$, $\tilde L$,
$\tilde u_R$ and $\tilde e_R$ are degenerate
(barring small $F$--term contributions) at the
scale of $\sqrt\xi$. We will return to this point in section 5.

It has been suggested in ref. \cite{dvali}, that in a supergravity theory
the squarks and the sleptons are expected to
receive contributions to their masses from the K\"ahler potential
through $F$--terms like
$\lambda\int d^4\theta(\bar\Phi_{45})({\bar\Phi_{45}})^\dagger
q_iq_i^\dagger/ M_{\rm Pl}^2$, where $\lambda={\cal O}(1)$.
Although these operators conserve all gauge symmetries,
in a string theory one still needs to ascertain whether they satisfy the 
string--selection rules; otherwise $\lambda$ would be small ($\le1/10$)
compared to unity. Deferring the study of this issue to a later work, 
we note that the contribution of these terms, if they are present,
to squark masses are given by
\beq
[\Delta m_{{\tilde q}_i}^2]_F\approx
{\lambda\vert\langle F({\bar\Phi}_{45})\rangle\vert^2\over{M^2_{Pl}}}
\approx {\lambda m^2{\hat\xi}\over{2M_{Pl}^2}}=\lambda m^2\epsilon/2
\label{Ftermtomsq}
\eeq
where $\epsilon={\hat\xi}/M_{pl}^2$. With ${\hat\xi}=\sqrt{3}\xi,$
and $\xi$ given by (\ref{dxi})
and ${\rm Tr}Q_A=72/\sqrt{3}$ \cite{top}, we expect $\epsilon\approx1/60$.
In general, the $F$--term contributions are not expected
to be universal, unless the K\"ahler potential possesses a certain symmetry
(see remarks below). Even then, and even if $\lambda\approx(1/2-1)$ (say),
these
$F$--term contributions are suppressed compared to the $D$--term contribution
(eq. (\ref{msqda})) by about a factor of (60--30),
for $\epsilon\approx1/60$.
Degeneracy to this extent suffices to account for
the smallness of at least the real part of the
$K^0-{\bar K}^0$ transition amplitude, if
$m_{\tilde d}\approx m_{\tilde s}\ge (700-1400){\rm GeV}$
\cite{DKS}\footnote{In quoting lower limits on squark--masses, we have
used a value for the product of mixing angles $(\cos\theta_d)
(\sin\theta_d)\approx(1/8-1/10)$, for the down quark--sector,
which seems reasonable.}.
Understanding the extreme smallness of the imaginary part
of the $K^0-{\bar K}^0$ amplitude, which we do not address here
would need additional considerations, based perhaps on symmetry
properties, which may explain why the relevant phase angle is
so small $\le 10^{-2}$\footnote{The problem of SUSY CP--violation,
in the context of models of SUSY--breaking as proposed here,
is discussed in a forthcoming paper by K.S. Babu and J.C. Pati,
where a natural explanation for the extreme smallness of
the $\epsilon$--parameter is given.}.

At this stage, the following property of the string solutions
under study is worth noting. We have observed that the family
permutation symmetry is exact at the level of the NAHE set in that 
it holds for the gauge interactions as well as for the super
and K\"ahler potentials. Even after the introduction of the
additional boundary condition basis vectors $(\alpha,\beta,\gamma)$
the permutation symmetry is still retained in the gauge
interactions, as well as in the cubic Yukawa interactions of the
quarks and the leptons with the Higgs fields and in the
$h_i{\bar h}_j{\bar\Phi}_{ij}$--terms of the superpotential $W$
(see eq. (\ref{superpot})). It is lost in $W$ only through (a)
${\cal O}(\Phi^3)$--terms, (b) terms involving Higgs and
the exotic fields but not quarks and leptons (these are not
shown in (\ref{superpot})), and (c) possibly some higher
dimensional terms. As a result, the effect of this loss
of the permutation symmetry on the quarks and leptons
and their superpartners is extremely mild in that it is
felt by them only through at most two--loop effects and higher
dimensional terms (whose contributions to the masses of the
$(d,s)$--squarks are less than or of order 200 MeV).
If the K\"ahler potential retains the family permutation symmetry
to the same extent as the superpotential $W$, which is plausible,
but which is an issue that needs to be examined, even the $F$--term
contributions given by (\ref{Ftermtomsq}) would be very nearly family
universal. In this case, and/or if $\lambda\le1/10$, the degree
of squark--degeneracy would be far better (in this case,
one would have $[({\tilde m}_i^2-{\tilde m}_j^2)/{\tilde m}^2]<<10^{-2}$)
than that indicated above. We defer the study of the K\"ahler
potential to later work. For the present, we will proceed by taking the 
squark degeneracy ratio to be no better than $1/30-1/60$, as obtained
above. 

The gauginos of the Standard Model gauge sector ({\it i.e.} gluinos,
winos etc.) could, in general,
receive masses through operators of the form:
$\lambda^\prime\int d^2\theta\Phi_{45}
{\bar\Phi}_{45}W_aW_a/M_{Pl}^2$ $(a=1,2,3)$ \cite{dvali} which yields:
\beq
m({\lambda_a})\approx \lambda^\prime
\langle F({\bar\Phi}_{45})\langle\Phi_{45}\rangle/
M_{Pl}^2\approx \lambda^\prime\epsilon m
\label{gauginomasses}
\eeq
where $\lambda^\prime\le{\cal O}(1)$. We see the hierarchy
\beq
[m^2(\tilde q_i)\approx Q_A^i(\sqrt{3}m^2/2)]~>~
[\Delta m_{\tilde q_i}^2\approx
\lambda\epsilon(m^2/2)]~>~ [m_{\lambda_a}^2\approx\lambda^\prime\epsilon^2m^2]
\label{sqm2}
\eeq

Because of this hierarchy, it is clear that if SUSY breaking
proceeds entirely through anomalous $U(1)$, the gluinos typically
would be rather light. From (\ref{gauginomasses}),
one obtains: $m_{\tilde g}\approx 2\lambda^\prime\epsilon m_{\tilde q}
\approx \lambda^\prime(20-60){\rm GeV}$, for $m_{\tilde q}\approx
(1-3){\rm TeV}$; this may be too light,
compared to the observed limit on $m_{\tilde g}$ of 130GeV,
unless $\lambda^\prime\ge2$
and $m_{\tilde q}\ge3{\rm TeV}$. To make matters worse, for string
solutions, as considered here, $\lambda^\prime$ vanishes at tree
level and can only arise through quantum loops; thus it is expected
to be small. This suggests that SUSY
breaking through anomalous $U(1)$, quite plausibly, is accompanied
by an additional source which provides the dominant contribution to
gluino masses $(\sim(1-~{\rm few})(100{\rm GeV}))$, while preserving
the squark--degeneracy, obtained through $U(1)_A$. We comment
on this possibility in section 6.

We should note that, for the sake of convenience, we have
evaluated the VEVs of $\Phi_{45}$, ${\bar\Phi}_{45}$, and
${\bar\Phi}_{45}^\prime$ and the auxiliary fields in the flat limit.
It has, however, been shown in Ref. \cite{dvali}, that the
inclusion of supergravity effects do not restore
supersymmetry, though they shift the VEVs of fields; {\it e.g.}
$\langle{\bar\Phi}_{45}\rangle$ acquires a non--zero value
which is typically bounded above by $\langle\Phi_{45}\rangle$.
Such shifts, however, do not alter the pattern of soft
masses and the hierarchy shown in (\ref{sqm2}).

Before discussing the origin of the mass term $m$
and certain phenomenological issues, we first discuss SUSY--breaking
in solution I (\cite{eu}).

$\underline{\hbox{{\bf SUSY breaking in solution I}:}}$

This case is more complex than the one presented
above, because it has six $U(1)'s$ (in contrast
to three relevant ones for solution II), four of which are non--universal,
and typically several fields (not merely two) must acquire VEVs
of order $\sqrt\xi$ to satisfy the $F$ and $D$ flat
conditions. The instructive example presented above prompts
us nevertheless to ask: (i) can one still find at least
a local minimum of the potential V which leads to nonzero
VEVs for the $D$--terms of only the family universal
charges -- i.e. $Q_A$ and $Q_\chi$ ?
(ii) If so, is that a global minimum ?
We find that the answer to the first question,
interestingly enough is in the affirmative,
and that to the second, though hard to assess in general, is
also found to be the same for the limited subset of
field--space, considered here.

Consider now a solution to the $D$ and $F$ flat conditions
(eqs. (\ref{dterms}--\ref{dxi}) for the case of solution I (ref. \cite{eu}),
which assigns non--zero VEVs of order $\sqrt\xi$
to the following set of fields:
\beq
\{ \Phi_{45},{\bar\Phi}_{13},{\bar\Phi}_3^-,{\bar\Phi}_1^+,
{\bar\Phi}_2^-,\xi_1\}\rangle={\cal O}(\sqrt{\xi})
\label{patternsol2}
\eeq
All other fields have zero VEVs at the scale $\sqrt{\xi}$.

The contributions of these fields to the $D$--terms of
the symmetries listed in eqs (\ref{u1a}) and (\ref{u12upsi}) are
given by (compare with eqs. (\ref{daka}--\ref{d274})).
\beqn
D_A &=& {1\over\sqrt{15}}[K_A-\sigma^2+
		4(\vert\bar\Phi_{45}\vert^2-
		\vert{\Phi}_{45}\vert^2)+{\hat\xi}]\label{sol1danom}\\
D_\psi &=& {1\over\sqrt{6}}[K_\psi-
			3\vert{\bar\Phi}_{13}\vert^2+(\vert\Phi_{45}\vert^2
			-\vert\bar\Phi_{45}\vert^2)]\label{sol1daka}\\
D_{12} &=& {1\over\sqrt{2}}[K_{12}-
		\vert{\bar\Phi}_{13}\vert^2-2\vert{\bar\Phi}_{12}\vert^2
	+{\sigma^2\over 3} -{\beta^2\over 3}-\delta^2]\label{sol1d12k12}\\
D_{45} &=& {1\over\sqrt{2}} [K_{45}+\vert\delta\vert^2]\label{sol1dzeta}\\
D_\zeta &=& {1\over\sqrt{6}}[K_\zeta+\beta^2]\label{d45dzeta}\\
D_\chi &=& {1\over\sqrt{15}}[K_\chi+2\sigma^2+ 2(\vert{\bar\Phi}_{45}\vert^2-
\vert\Phi_{45}\vert^2)]
\label{dchi}
\eeqn
Here ${\hat\xi}\equiv\sqrt{15}\xi$.
As before, the K--terms contain fields like squarks, sleptons
and Higgs--bosons which have zero VEVs.
Anticipating that a field like $\bar\Phi_{12}$ (or $\Phi_{23}$)
which is charged under $U_{12}$ and possibly $U_\psi$,
but not the other $U(1)'s$, may need to acquire
a VEV of order $m<< \sqrt{\xi}$, in the presence of a
mass--term $m$, we have exhibited its contribution.
The combinations $\sigma$, $\beta$ and $\delta$ are defined by:
\beqn
&& \sigma^2\equiv\vert\Phi^+_1\vert^2+\vert{\bar\Phi}_2^-\vert^2+
\vert{\bar\Phi}^-_3\vert^2\\
&& \beta^2\equiv\vert\Phi^+_1\vert^2+\vert{\bar\Phi}_2^-\vert^2-
2\vert{\bar\Phi}^-_3\vert^2\\
&& \delta^2\equiv\vert\Phi^+_1\vert^2-\vert{\bar\Phi}_2^-\vert^2
\label{sigzedel}
\eeqn
It may be verified that all the $F$-- and $D$--flat constraints
($F_i=D_A=D_\alpha=0$) are satisfied for the $m=0$ cubic--level
superpotential, with $\vert\langle\Phi_{45}\rangle_0\vert^2=
{\hat\xi}/5$, and
$\vert\langle{\bar\Phi}_{13}\rangle_0\vert^2=
 \vert\langle\Phi_1^+\rangle_0\vert^2=
 \vert\langle{\bar\Phi}_2^-\rangle_0\vert^2=
\vert\langle{\bar\Phi}_3^-\rangle_0\vert^2={\hat\xi}/15$, {\it i.e.}
\beqn
 \vert\langle\Phi_{45}\rangle_0\vert^2=\vert\langle\sigma\rangle_0\vert^2=
3\vert\langle{\bar\Phi}_{13}\rangle_0\vert^2={{\hat\xi}\over5}\\
 \langle\beta^2\rangle_0=\langle\delta^2\rangle_0=
 \langle{\bar\Phi}_{12}\rangle_0={0}
\eeqn
All other VEVs are zero. As before, the subscript zero signifies that
the VEVs are obtained in the limit of zero--mass for
all the fields. The VEV of the singlet $\xi_1$ is not determined
by the $F$ and $D$ flat constraints (at least at the cubic level
superpotential). Independent phenomenological
considerations including quark--lepton masses
and mixings, however imply that the VEV $\langle\xi_1\rangle$ must
be of order $\sqrt{\xi}\sim{\cal O}(g^2M_{st}/4\pi)$ \cite{nrt}.
In the absence of a complete solution to the vacuum selection
in string theory, we will proceed by imposing this choice.

Allowing for a mass term $m$ as in (\ref{massterm}), and extremizing the
potential
\beq
V={g^2\over2}\sum_\alpha D_\alpha^2+m^2(\vert\Phi_{45}\vert^2
+\vert{\bar\Phi}_{45}\vert^2)
\eeq
with respect to $\Phi_{45}$, $\bar\Phi_{13}$, $\sigma$ and
${\bar\Phi}_{12}$, respectively we obtain (compare with
eqs. (\ref{sol2vevp45})--(\ref{sol2vevp45b})):
\beqn
 {{\partial V}\over{\partial\Phi_{45}}} &=& 0\Rightarrow
{{(\Phi_{45})^\dagger}}[4(D_A+{D_\chi\over2}-
	{\sqrt{15}\over4}{m^{\prime^2}})-{\sqrt{5\over2}}
					D_\psi]=0\label{sol1vevp45}\\
 {{\partial V}\over{\partial{\bar\Phi}_{13}}} &=& 0 \Rightarrow
{{({\bar\Phi}_{13})^\dagger}}
		 [D_{12}+\sqrt{3} D_\psi]=0\label{sol1vevp13}\\
 {{\partial V}\over{\partial\sigma}} &=& 0 \Rightarrow
{\sigma}
		[(D_A-2D_\chi)-\sqrt{5\over6}
		{D_{12}}]=0\label{sol1vevsig}\\
 {{\partial V}\over{\partial{\bar\Phi}_{12}}} &=& 0 \Rightarrow
{{({\bar\Phi}_{12})^\dagger}} (D_{12})=0\label{sol1vevp12}
\eeqn
Here $m^{\prime^2}\equiv{m^2/g^2}$.
variations with respect to $\beta$ and $\delta$
are not exhibited because these can be satisfied
consistently by preserving their zero
mass values: $\beta=\beta_0=0$ and $\delta=\delta_0=0$.
Thus, from eqs. (\ref{sol1dzeta}) and (\ref{d45dzeta}),
$D_{45}=D_\zeta=0$. As in the previous
example, we see from eq. (\ref{sol1vevp45}) that $\langle D_A\rangle$,
$\langle D_\chi\rangle$ and/or $\langle D_\psi\rangle$ must be
of order $(m^2)$, and thus SUSY is broken.

Unlike the previous example, however, where
$\Phi_{45}^\prime\ne0$ uniquely led (via eq. (\ref{sol2vevs})) to
$D_\psi=0$, we see that eq. (\ref{sol1vevp13}), can be satisfied,
given ${\bar\Phi}_{13}\ne0$, by choosing either (a)
$D_{12}=D_\psi=0$, or (b) $D_{12}=-\sqrt{3}D_\psi={\cal O}
(m^2)\ne0$. In short,
the solution for the $D$'s do not appear to be unique.
Case (a) would, of course be phenomenologically
preferable, because it would lead to family universal
squark masses. We consider these two cases by turn
and show that minimization
of the potential in fact favors case (a) over case (b).

$\underline{\hbox{Case (a) }D_{12}=D_\psi=0}$

Given $D_{12}=0$, eq. (\ref{sol1vevp12}) can be satisfied
by choosing either $\bar\Phi_{12}=0$ or
${\bar\Phi}_{12}={\cal O}(m)\ne0$.
We will see that internal consistency
will fix $\bar\Phi_{12}={\cal O}(m)\ne0$, if $D_{12}=0$.

Given $D_{12}=D_{\psi}=0$, eqs. (\ref{sol1vevp45})
and (\ref{sol1vevsig})
imply: (i) $D_A+D_\chi/2=\sqrt{15}m^{\prime^2}/4$ and (ii)
$D_A=2D_\chi$, which in turn imply:
\beq
D_A=\sqrt{3\over5}m^{\prime^2}~~~~;~~~~
D_\chi=\sqrt{3\over5}{m^{\prime^2}\over{2}}~,
\label{a1}
\eeq
Now, given $\beta=\delta=0$, looking at the compositions
of the $D$--terms (eqs. (\ref{sol1daka}--\ref{dchi})),
$D_\psi=0$ and $D_{12}=0$ imply:
\beqn
\vert{\bar\Phi}_{13}\vert^2 &=& \vert{\Phi}_{45}\vert^2/3\label{a2}\\
\sigma^2 &=& \vert\Phi_{45}\vert^2+6\vert{\bar\Phi}_{12}\vert^2\label{a3}
\eeqn
Substituting (\ref{a3}) in $D_\chi$ (see eq. (\ref{dchi})),
and putting $D_\chi=\sqrt{3/5}(m^{\prime^2}/2)$ (see (\ref{a1})), we get
\beq
\vert{\bar\Phi}_{12}\vert^2={m^{\prime^2}\over{8}}
\label{a4}
\eeq
Thus, internal consistency for case (a)
($D_{12}=D_\psi=0$)
imply that a field like $\bar\Phi_{12}$
(alternatively $\Phi_{23}$ will also be adequate),
which had a zero VEV to begin with ({\it i.e.} for $m=0$),
must acquire a non-zero VEV of order $m$.

Solving for the nonzero VEVs, and collecting the results, we obtain
\beqn
&& \vert\langle\Phi_{45}\rangle\vert^2={{\hat\xi}\over5}-
				{3\over4}{m^{\prime^2}}~~;~~
   \vert\langle{\bar\Phi}_{13}\rangle\vert^2={{\hat\xi}\over{15}}-
			{m^{\prime^2}\over{4}}~~;~~
   \langle\sigma^2\rangle={{\hat\xi}\over5}\\
&& \langle{\bar\Phi}_{12}\rangle^2={m^{\prime^2}\over{8}}~~;~~\\
&& \langle{\bar\Phi}_{45}\rangle^2=\langle\beta^2\rangle=
   \langle\delta^2\rangle=0\\
&& \langle D_A\rangle=\sqrt{3\over5}{m^{\prime^2}}~~;~~
   \langle D_\chi\rangle=\sqrt{3\over5}{m^{\prime^2}\over{2}}~~;~~\\
&&  D_\psi=D_{12}=D_\zeta=D_{45}=0\\
&& \langle F(\Phi_{45})\rangle=\langle F({\bar\Phi}_{13})\rangle=
   \langle F(\sigma)\rangle=\langle F(\zeta)\rangle=
   \langle F(\delta)\rangle=0\\
&& \langle F({\bar\Phi}_{45})\rangle=m\sqrt{{{\hat\xi}\over5}-
			{3\over4}{m^{\prime^2}}}
\label{278shiftedvevs}
\eeqn
It may be verified that the solution presented
above is in fact a minimum of $V$.
{\it We see that
there exists at least a local minimum
for which the $D$--terms of only the universal
charges $Q_A$ and $Q_\chi$ are non--zero}. This
solution thus has the desirable feature that the
$D$--term contributions to the squark--masses,
which dominate over $F$--term contributions,
are family universal. The rest of the phenomenological
discussions ({\it i.e.} the gaugino
masses and the hierarchy) are qualitatively
the same as in solution II (see eqs \ref{msqda}--\ref{sqm2}).
To be specific, now ${\rm Tr}Q_A=180/\sqrt{15}$,
but ${\hat\xi}=\sqrt{15}\xi$,
so $\epsilon\equiv{\hat\xi}/M_{Pl}^2\approx1/25$.
The contributions of $D_A$ and $D_\chi$
are given by:
$[m^2_{\tilde d, \tilde s, \tilde b}]_{D_A,D_\chi}=
g^2[Q_AD_\chi+Q_\chi D_\chi]=m^2/4,$
where we have put $\sqrt{15}Q_A=3/2$ and $\sqrt{15}Q_\chi=-1/2$
(see table 1) and $D_A=\sqrt{3/5}m^{\prime^2}=2D_\chi$.
The $F$--term contributions (see eq. (\ref{Ftermtomsq})),
which may in general be non--universal, are given by:
$[m^2_{\tilde d, \tilde s, \tilde b}]_{F}\approx
\lambda\vert F({\bar\Phi}_{45})\vert^2/M_{Pl}^2\approx
\lambda(m^2/5)\epsilon,$ which are thus suppressed
by about a factor of (1/60--1/30) (for $\lambda\approx(1/2-1)$) compared
to the universal $D$--term contributions of
$g^2[Q_AD_\chi+Q_\chi D_\chi]=m^2/4.$
As for solution II, this is compatible with the constraints from the
real part of the $K^0-{\bar K}^0$ amplitude, if
$m_{{\tilde d},{\tilde s}}\ge(700-1400){\rm GeV}$.

$\underline{\hbox{Case (b): }D_{12}={\cal O}(m^2)\ne0}$

In this case, eqs. (\ref{sol1vevp12}) and (\ref{sol1vevp13})
respectively imply:
\beq
{\bar\Phi}_{12}=0, ~{\rm and}~ D_\psi=-D_{12}/\sqrt{3}
\label{casebb1}
\eeq

substituting (\ref{casebb1}) into (\ref{sol1vevp45})
and using (\ref{sol1vevsig}),
we get
\beq
D_A=\sqrt{3\over5}{m^{\prime^2}}~~~~;~~~~
D_\chi={1\over2}(\sqrt{3\over5}{m^{\prime^2}}-
			\sqrt{5\over6}{D_{12}})
\label{casebb2}
\eeq
Given $\bar\Phi_{12}=\beta=\delta=0$, the expressions
for $D_A$, $D_{12}$ and $D_\psi$ and $D_\chi$ given in
eqs. (\ref{sol1danom}--\ref{dchi})
respectively yield:
\beqn
&& 4 \vert\Phi_{45}\vert^2+\sigma^2={\hat\xi}-3{m^{\prime^2}}\label{casebA}\\
&& -\vert{\bar\Phi}_{13}\vert^2+
			{\sigma^2\over3}=\sqrt{2}D_{12}\label{casebB}\\
&& -3 \vert{\bar\Phi}_{13}\vert^2+
		\vert{\Phi_{45}}\vert^2=-\sqrt{2}{D_{12}}\label{casebC}\\
&& \sigma^2-\vert\Phi_{45}\vert^2={3\over4}m^{\prime^2}-
			{5\over{4\sqrt{2}}}D_{12}
\label{casebE}
\eeqn
Combining (\ref{casebB}) and (\ref{casebC}), we get
\beq
\vert\Phi_{45}\vert^2-\sigma^2=-4\sqrt{2}D_{12}\label{casebD}
\eeq
Combining (\ref{casebA}--\ref{casebD}), one obtains
\beqn
&& \vert\Phi_{45}\vert^2={{\hat\xi}\over5}-{3\over4}{m^{\prime^2}}+
			{{1}\over{4\sqrt{2}}}D_{12}\label{casebb3}\\
&& \sigma^2={{\hat\xi}\over5}-{{1}\over{\sqrt2}}D_{12}\label{casebb4}\\
&& \vert{\bar\Phi}_{13}\vert^2={{\hat\xi}\over{15}}-
		{m^{\prime^2}\over4}+
		{3\over{4\sqrt{2}}}{D_{12}}
\label{casebb5}
\eeqn
Comparing (\ref{casebE}) and (\ref{casebD}), we get
\beq
D_{12}={{3\sqrt{2}}\over{37}}m^{\prime^2}
\label{casebb6}
\eeq

One can verify that the solution for the VEVs presented above
again corresponds to a minimum of $V$.
To compare the minimum obtained in case (a)
with that of case (b) we need to study the variation
of $V$ with respect to $D_{12}$. To do so
we evaluate the potential at the string unification scale,
where the couplings of the $U(1)$'s are unified.
\beq
V={g^2\over2}(D_A^2+D_\chi^2+D_\psi^2+D_{12}^2)+m^2(\vert\Phi_{45}\vert^2+
{\bar\Phi}_{45}^2)
\eeq
Substituting for $D_A$, $D_\chi$, $D_\psi$, and $\vert\Phi_{45}\vert^2$
from eqs. (\ref{casebb2}), (\ref{casebb1}), and (\ref{casebb3}),
we get\footnote{The question of why the vacuum energy (cosmological
constant) is so small or zero
of course remains unanswered, as it is in all
other analogous approaches leading to SUSY--breaking.}
\beq
{V\over{(g^2/2)}}={2\over5}{\hat\xi}m^{\prime^2}+{\cal O}(m^{\prime^4})+
{{13}\over{24}}D_{12}^2+m^{\prime^2}(-{1\over{4\sqrt2}}+
			{1\over{4\sqrt2}})D_{12}
\label{casebput}
\eeq
We see that the coefficient of the linear term
in $D_{12}$ cancels owing to contributions from $D_\chi^2$
and $\vert\Phi_{45}\vert^2$, and thus $V$ would increase
for $D_{12}\ne0$. This shows that the minimum of $V$
corresponding to case (a), with $D_{12}=D_\psi=0$,
is preferred over that of case (b), with
$D_{12}=-\sqrt{3}D_\psi\ne0$.
This in turn means that even for the more realistic, though complicated,
case of solution I, there exist viable solutions for the pattern
of VEVs for which only the family--universal contributions
to squark masses, arising through $D_A$ and $D_\chi$,
survive, but the non--universal $D$--term contributions,
associated with the family dependent $U(1)$'s, vanish
owing to minimization of the potential. While this
result may not hold in general, it is remarkable
that it does for the solutions considered here, which are viable.
The conditions for emergence of this result, which are worth
studying, will thus provide an important new guideline for
selecting the string solutions, and the associated patterns of
zeroth order VEVs which satisfy the $F$ and $D$--flat conditions.

The degeneracy in squark--masses, obtained as above at the string--unification
scale, would of course be affected, as would be the ratios of
the various gauge and Yukawa couplings, when they are extrapolated
to low energies through the use of the renormalization
group equations. This would not, however,
have a significant effect at least on the degeneracy
of the squarks of the first two families, which is relevant to
the $K_0-{\bar K}_0$ transition\footnote{We should also add
that even though $\langle D_{12}\rangle$ and $\langle D_\psi\rangle$
vanish at the level considered above, the VEVs of Higgs
fields (like $H_u$) of electroweak scale will
still induce a non--vanishing $\langle D_{12}\rangle\approx
Q_{12}(H_u)\vert\langle H_u\rangle\vert^2=(1/\sqrt{2})\vert\langle
H_u\rangle\vert^2$, and likewise a non--vanishing $\langle D_\psi\rangle$.
This leads to a mass splitting $\vert\delta{m^2_{\tilde d}}-
\delta{m^2_{\tilde s}}\vert\approx(g_2^2/2)\vert
(Q_{12}^{\tilde d}-Q_{12}^{\tilde s})\vert\langle D_{12}\rangle
\approx(g_2^2/2)(1/(2\sqrt{2}))(1/\sqrt{2})\vert\langle H_u\rangle\vert^2
\approx(50{\rm GeV})^2\le(1/186){\tilde m}^2$,
for ${\tilde m}\ge700{\rm GeV}$, where we have put
$\langle H_u\rangle\sim 200{\rm GeV}$.
As discussed in the text, lack of squark degeneracy to this extent
is of course compatible with the constraint of the
real part of the $K_0-{\bar K}_0$ amplitude.
We thank K.S. Babu for raising this point.}.

$\underline{\hbox{{\bf sign of $\langle D_A\rangle$ - Contribution to
Squares of Scalar Masses}:}}$

Before discussing the origin of the mass--term $m$,
one special property of both solutions I and II is worth
noting. Given that the overall sign of $U(1)_A$ is chosen
such that ${\rm Tr}Q_A$ is positive, the signs of the
anomalous charges of all the fields are fixed. For instance, if the
sign of $Q_A$ for squarks and/or sleptons in any solution happened
to be negative, it must of course be discarded because the
corresponding $D_A$ would lead to negative contributions to the
(mass)$^2$ of these fields (see Eq. (\ref{msqda}) and thereby
to a breaking of $SU(3)$--color and/or electric charge.
As may be seen from tables 1 and 4, it is indeed remarkable
that {\it all the squarks and the sleptons have positive $Q_A$
for both solutions I and II}.

One must still ensure that none of the other fields
carrying color and/or electric charge acquire net
negative (mass)$^2$. Note first of all that
the hidden sector fields $V_i$, ${\bar V}_i$
as well as $T_i$ and ${\bar T}_i$ (which are of course
standard model singlets) have positive $Q_A$.
The fields of possible concern for solution I
are the conjugate pairs $(D_{45},{\bar D}_{45})$
and $(H_{21},H_{22})$ which carry color.
Clearly one member of each such pair would have positive
$Q_A$ but the other member would have negative $Q_A$,
as do $D_{45}$ and $H_{21}$. Thus $\langle D_A\rangle$
would give negative (mass)$^2$ to $D_{45}$ and $H_{21}$.
We have, however, checked that higher dimensional operators
for solution I as well as solution II give sufficient
positive contribution to the (mass)$^2$ of each member
of these conjugate pairs, by utilizing the VEVs of standard model
singlets of order $\sqrt{\xi}$ (as in Eq. (\ref{patternsol2}))
and hidden sector condensates. This more than compensates
for the negative contribution of $\langle D_A\rangle$. Specifically,
for solution I, one obtains the operators $H_{21}H_{22}\xi_1$
and $D_{45}{\bar D}_{45}H_{19}H_{20}\xi_1^3$ at $N=3$ and $N=7$
respectively in the superpotential $W$. With $\xi_1\sim\sqrt\xi$
(see Eq. (\ref{patternsol2})), and assuming that the
$(H_{21}H_{20})$ -- pair condenses due to the $SU(5)_H$
force which confines at a scale of $10^{13}-10^{14}$GeV \cite{fh2},
these operators provide positive contributions to (mass)$^2$
of $H_{21}$ and $H_{22}$ as well as of $D_{45}$ and ${\bar D}_{45}$,
that far exceed the negative contributions of $\langle D_A\rangle$,
which are of order $m^2$ [it is worth mentioning in advance
that $m$ itself is induced only at $N=8$ by utilizing condensate
of the same type as above (see next section)].

Now each member of a conjugate pair of Higgs doublets like
$(h_i,{\bar h}_i)$ would also get negative contribution
to its (mass)$^2\sim(1{\rm TeV})^2$ through $\langle D_A\rangle$.
As mentioned in section 3, the (mass)$^2$ matrix of the Higgs--sector,
including contributions from the string generated higher dimensional
operators has many entries. This matrix has been analyzed in detail
in Ref. \cite{nrt}, which showed that only one pair of doublets
remains light, while the others acquire heavy or medium heavy masses.
While a reanalysis of the Higgs mass--matrix including
the $\langle D_A\rangle$--contributions of order $(1{\rm TeV})^2$
deserves study in a separate work, it is clear that the latter
contribution will affect only the light Higgs spectrum. [in general,
it is possible that such a light Higgs may even acquire a VEV
of order $1{\rm TeV}$ due to the $\langle D_A\rangle$--contribution
at a high scale; this by itself need not, however, be objectionable].

In summary, we note that the higher dimensional operators
could not in any case have given masses to standard model
{\it non--singlet chiral} fields like squarks and sleptons
by using VEVs of only standard model singlet fields like those
in Eq. (\ref{patternsol2}) and hidden sector condensates.
It is thus fortunate that these fields carry positive anomalous
charges and thereby receive only positive contributions
to their (mass)$^2$ from $\langle D_A\rangle$ for both
solutions I and II. On the other hand, higher dimensional
operators can and do contribute positively to
the (mass)$^2$ of fields belonging to conjugate pairs,
and more than compensate for negative
contributions from $\langle D_A\rangle$ to such fields.

\setcounter{footnote}{0}
\section{Generating the mass term}

We now show how the mass term $m$ can arise through higher dimensional
operators using hidden sector condensates. Consider first solution I
(ref.\cite{eu}). The allowed operators up to $N=8$ are
listed below:

at order N=5,
\beqn
		  && V_2{\bar V_2}\Phi_{45}\Phi_2^-\xi_1,{\hskip 2cm}
                     V_1{\bar V_1}\Phi_{45}{\bar\Phi}_1^+\xi_2,\nonumber\\
                  && T_2{\bar T_2}\Phi_{45}\Phi_2^+\xi_1,{\hskip 2cm}
                     T_1{\bar T_1}\Phi_{45}{\bar\Phi}_1^-\xi_2,
\label{neq5}
\eeqn
at order N=7,
\beqn
		  && T_2{\bar T_3}V_3{\bar V_2}\Phi_{45}\Phi_{45}
					{\bar\Phi}_{13},\nonumber\\
		  && T_1{\bar T_2}V_1{\bar V_2}\Phi_{45}
					\Phi_{45}\xi_1,\nonumber\\
		  && T_2{\bar T_1}V_1{\bar V_2}\Phi_{45}\Phi_{45}\xi_2,
\label{neq7}
\eeqn
\beqn
&& V_2{\bar V_2}\Phi_{45}\Phi_2^-\xi_1
   [({{\partial W_3}\over{\partial\xi}_3})+\xi_i\xi_i+
   \Phi_{13}{\bar\Phi}_{13}+\Phi_{23}{\bar\Phi}_{23})],\nonumber\\
&& V_1{\bar V_1}\Phi_{45}{\bar\Phi}_1^+\xi_2
   [({{\partial W_3}\over{\partial\xi}_3})+\xi_i\xi_i+
   \Phi_{13}{\bar\Phi}_{13}+\Phi_{23}{\bar\Phi}_{23})],\nonumber\\
&& V_2{\bar V_2}\Phi_{45}{\bar\Phi}_2^+\xi_1
   ({{\partial W_3}\over{\partial\Phi}_{12}}),\nonumber\\
&& V_1{\bar V_1}\Phi_{45}{\Phi}_1^-\xi_2
   ({{\partial W_3}\over{\partial{\bar\Phi}_{12}}}),
\label{neq7p}
\eeqn
and at order N=8,
\beqn
&& {\bar T_2} T_3 V_3{\bar V_2}\Phi_{45}\Phi_{45}{\bar\Phi}_{13}
   \xi_1, \nonumber\\
&&  T_1 {\bar T_3} V_1{\bar V_3}\Phi_{45}\Phi_{45}{\bar\Phi}_{23}
		\xi_2, \nonumber\\
&&  T_2 {\bar T_3} V_2{\bar V_3}\Phi_{45}\Phi_{45}{\bar\Phi}_{13}
		\xi_1, \nonumber\\
&&  T_3 {\bar T_1} V_3{\bar V_1}\Phi_{45}\Phi_{45}{\bar\Phi}_{23}
    \xi_2.
\label{neq8}
\eeqn
\beqn
&&{\Phi}_{45}{\bar\Phi}_{45}T_3{\bar T}_3\Phi_{45}
			{\bar\Phi}_{13}{\bar\Phi}_3^-\xi_1,\nonumber\\
&&{\Phi}_{45}{\bar\Phi}_{45}T_3{\bar T}_3\Phi_{45}
			{\bar\Phi}_{13}{\bar\Phi}_3^-\xi_2,
\label{n8}
\eeqn
We see from (\ref{neq5}), that given the pattern of
VEVs listed in eq. (\ref{patternsol2}), no mass term can arise
at $N=5$. At $N=7$ bilinear mass--terms like
$\Phi_{45}^2$ could arise, only if non-diagonal hidden sector
condensates like $\langle T_2{\bar T}_3\rangle$
as well as $\langle V_3{\bar V}_2\rangle$ could form. For ordinary
QCD, such condensates (like $\langle{\bar d}s\rangle$) are forbidden.
Even if they do form, the magnitude of such mass--terms is
of order $(\Lambda_5/M_{\rm st})^2(\Lambda_3/M_{\rm st})^2M^\prime$,
where $\Lambda_5$ and $\Lambda_3$ represent the $SU(5)_H$ and
$SU(3)_H$ confinement scales, respectively, and $M^\prime$
is of order $\sqrt\xi$,
representing the scale of the VEVs of the singlet $\Phi$-fields
in eq. (\ref{patternsol2}).
Taking typical values of $\Lambda_5\sim 10^{13}-10^{14}{\rm GeV}$
and $\Lambda_3\sim 10^{8}-10^{10}{\rm GeV}$, which are suggested by
renormalization group analysis for solution I \cite{fh2}.
With
$M_{\rm st}\sim 10^{18}{\rm GeV}$ and $M^\prime\sim(1/3-1)\times
10^{17}{\rm GeV}$,
such mass--terms are $\le10^{-9}{\rm GeV}$, and are thus
insignificant for
phenomenological purposes.
Assuming that at least the diagonal condensates in the hidden
sector, like $\langle T_3{\bar T_3}\rangle$ form, the only
relevant mass--term is given by the $N=8$ term shown
in Eq. (\ref{n8}). This yields
a mass term $m\Phi_{45}{\bar\Phi}_{45}$ which is neutral
with respect to all charges and is of order :
\beq
m(\Phi_{45})=m({\bar\Phi}_{45})\sim\left({{\Lambda_5}\over{M_{st}}}\right)^2
\left({{M^\prime}\over{M_{st}}}\right)^3 M^\prime
\label{ordermp45}
\eeq
It is remarkable that $m$ receives contributions
only at $N\ge8$. Since $\Lambda_5$ is 4 to 5 orders
of magnitude smaller, and $M^\prime$ is about 10 to 30
times smaller than $M_{st}$, it is clear that the SUSY mass
splitting $m$ is naturally strongly suppressed compared to $M_{st}$.
As regard its numerical value, for values of
$\Lambda_5$ and $M^\prime$ lying in the range mentioned
above, {\it i.e.} $(\Lambda_5/M_{st})^2\sim10^{-8}-10^{-10}$,
$(M^\prime/M_{st})^3\sim10^{-4}$ and $M^\prime\sim(1/2)(10^{17}{\rm GeV})$,
say -- which are most plausible, we get :
\beq
m~\sim~({1\over2}-50) {\rm TeV}.
\label{msusy}
\eeq
Since $m$ represents the scale of supersymmetry breaking and thus
the mass scale of the Higgs scalars, and in turn of $m_W$,
we see that the string solutions under consideration do
explain why the electroweak scale is so much smaller than the
string scale.

\setcounter{footnote}{0}
\section{Some phenomenological aspects}

An interesting phenomenological distinction, first between the
two solutions I and II, considered above, is worth noting.
Although the anomalous charge $Q_A$ for solution I \cite{eu},
given by eq. (\ref{u1a}), is family -- universal, it distinguishes
between different members of a family (see table 1), and therefore
leads to intra--family mass splittings among the scalars.
Including contributions from the leading $D_A$ term
only, these are represented, by the following relative values,
at the scale of $\sqrt\xi$, for any given family :
\beq
\left[m^2({\tilde Q}_L):m^2({\tilde u}_R):m^2({\tilde d}_R):
m^2({\tilde L}):m^2({\tilde e}_R)\right]_{D_A}=~3:1:3:1:1~~~
{\rm soln~I}
\label{ratiosol1}
\eeq

For solution II \cite{top}, on the other hand,
$Q_A$ given by eq. (\ref{sol2ua}), is the same for all members
of a family. As a result, in so far as the leading contributions
from the $D_A$--term, one obtains intra--family universal
scalar masses at the scale $\sqrt\xi$, which are given by :
\beq
\left[m^2({\tilde Q}_L):m^2({\tilde u}_R):m^2({\tilde d}_R):
m^2({\tilde L}):m^2({\tilde e}_R)\right]_{D_A}=~1:1:1:1:1~~~
{\rm soln~II}
\label{ratiosol2}
\eeq
Thus, eventually empirical study of the squark spectrum can
in fact distinguish between the two string solutions I and II.

It is also worth noting that both string solutions I and II
lead to approximately universal scalar masses (at the scale of $\sqrt\xi$)
{\it for all three families}. At the same time,
owing to spontaneously induced asymmetric Higgs mass spectrum (see
discussion in section 3) they lead to hierarchical fermion masses \cite{nrt}.
By contrast, the model of ref. \cite{dvali} assumes that
$U(1)_A$ couples universally only to the first
two families and thus predicts heavier squark masses
($\sim 5 {\rm TeV}$) for the first two
families and lighter mass $(\sim 500 {\rm GeV})$ for the stop,
while the gauginos are lighter still $(\sim50-100{\rm GeV})$.
It has been pointed out in ref. \cite{murayama},
however that models of this class \cite{dvali,mohapatra} with two--family
universality (of the anomalous $U(1)$) typically lead to color
and electric charge--breaking, assuming that the spectrum of
the type noted above is generated near the Planck or the GUT--scale.
This is because contributions from two--loop renormalization group
evolution to the scalar masses contain terms which
are proportional to the larger squark (mass)$^2$ of
the first two families, and are negative. This
negative contribution turns the initially smaller positive
stop (mass)$^2\sim(500{\rm GeV})^2$ to negative values
at the TeV--scale and thereby induces color and charge--breaking.
Models with three--family universality (of the anomalous $U(1)$),
as discussed here, do not however face this problem
because the squarks of all three families are nearly degenerate,
with only moderately heavy masses. Several considerations suggest
that they should have a mass of about 1 TeV, within a factor of
two, either way, at the electroweak scale.
This reduces the RGE--induced negative contribution to the
squark (mass)$^2$ by about an order of magnitude, while increasing
the initial positive value of $m_{\tilde t}^2$ at the GUT
or the string scale, compared to the case of two--family universality.
It thereby eliminates the problem of color and charge--breaking.
Thus it seems that phenomenological considerations favor
three--family universality, for SUSY--breaking through anomalous $U(1)$.
It is intriguing that string--solutions of the type considered
here yield precisely that.

\setcounter{footnote}{0}
\section{Remarks on Supersymmetry Breaking Through Anomalous $U(1)$}

Before concluding the following remarks are in order.

$\underline{\hbox{{\bf (1) Desirability and Origin of Family-Permutation
Symmetry}:}}$
If supersymmetry breaking occurs entirely
or dominantly through an anomalous $U(1)$, as noted
in the last section, the need to avoid color and charge--breaking
suggests that the {\it $U(1)_A$ must be universal
with respect to all three families}.
At the same time, the hierarchical masses and mixings
of the three families suggest that there ought to exist
{\it flavor or horizontal gauge symmetries}, beyond
GUTs in the underlying theory, which distinguish
between the families and are ultimately responsible
for the hierarchy in their masses.
The virtues of flavor symmetries in the string context,
(like $U_1$ to $U_6$ for solution I and $U_1$ to $U_5$
for solution II) in this regard has been noted in previous
works \cite{u1sfmh,nrt},
and in the non--string context by several authors \cite{u1ffmh}.
Furthermore, these flavor symmetries have been shown to play
a crucial role in addressing certain
naturalness problems of supersymmetry, such as the enormous
suppressions of (a) the $d=4$ and $d=5$
rapid proton--decay operators \cite{custodial,jp},
(b) $\nu_L-{\tilde{\rm H}}$ mixing mass \cite{nhmix},
and (c) the mass--term $m$ of relevant fields which triggers
SUSY--breaking(see section 4).
We suspect that they are also responsible for the desired
suppression of the $\mu$--parameter.

As alluded to above, such family--dependent flavor symmetries,
which are clearly absent in GUTs, do in fact emerge quite generically
in string theory ({\it e.g.} from an underlying $SO(44)$ in the
free fermionic construction). Now, typically, at least a subset
of these family--dependent $U(1)'s$ would appear to be anomalous
in a general basis (compare with $U_1$ to $U_6$ for solution I
and $U_1$ to $U_3$ for solution II); these can be grouped
to give anomaly--free combinations
(like $U_\psi,~U_{12},~U_\zeta,~U_{45}~$ and $U_\chi$ in solution
I (see eqs. (\ref{u12upsi}--\ref{uchi}));
except for one unique combination that remains anomalous and gives
the $U(1)_A$ (see eq. (\ref{u1a})). Similar situations
arise in all other semi--realistic string--derived models
which exist to date; see {\it e.g.} Refs. \cite{revamp,fny,alr,price,eu,top,
lykken,ibanez,kakushadze}.

The question then arises how can there be these flavor--symmetries,
which distinguish between the families and are
thus family  -- non-universal, and yet there be an anomalous
$U(1)_A$, arising from linear combinations of the same
flavor--symmetries, which is family universal ?
The only way, it appears to us, is that the flavor
symmetries, although family--dependent, must still
respect the {\it permutation symmetry} (mentioned in secs 1 and 2)
with respect to all three families. In this case $U(1)_A$ would
automatically be family universal, as borne out by the examples
of eqs. (\ref{u1a}) and (\ref{sol2u1com}).

Thus, SUSY--breaking through $U(1)_A$, together
with the presence of flavor symmetries, seem to suggest
the need for the stated permutation symmetry.
As stated in section 2, such a symmetry is in fact
an internal property of at least the NAHE set of
boundary condition basis vectors $\{{\bf1}, S,b_1,b_2,b_3\}$,
for which the cyclic permutation symmetry corresponds
simply to the symmetry between the three
twisted sectors of the $Z_2\times Z_2$ orbifold, which
arise from the sectors $b_1$, $b_2$ and $b_3$, respectively.
In view of the importance of the permutation symmetry, as noted above,
it would be interesting to know whether such a symmetry could
arise without utilizing the NAHE set.
While it is premature to ascertain the answer to this question
at present, we note that
there do in fact exist three--generation string solutions
based on the free fermionic construction which utilize only
a subset $\{{\bf1},S,b_1,b_2\}$ of the NAHE set of basis vectors
(see {\it e.g.} Ref. \cite{lykken}); these, however, do not possess
the cyclic permutation symmetry.

It is also worth noting that while the NAHE set (by itself)
yields the permutation symmetry, it of course does not guarantee
that the symmetry will be retained in the presence of additional
boundary condition basis vectors, which are needed to reduce
the number of generations from 48 to 3. As noted before,
all four solutions exhibited respectively in Refs.
\cite{revamp}, \cite{alr}, \cite{eu} and \cite{top}, utilize the NAHE
set; but only the last two retain the permutation symmetry,
while the first two do not. Thus, string solutions of the
type obtained in Refs. \cite{eu} and \cite{top} appear
to be particularly suited to break supersymmetry
through an anomalous $U(1)$, while providing the
squark degeneracy.

$\underline{\hbox{{\bf (2) A Scenario of Combined Anomalous $U(1)$--Dilaton
SUSY breaking}:}}$
It has been noted in section 3 (see discussion following eq. (\ref{sqm2}))
that if SUSY--breaking proceeds entirely through a family--universal
$U(1)_A$, it would lead to the desired squark degeneracy,
but it is likely to lead to unacceptably light gluinos. At this
point, two apparently unrelated issues, both associated with
the dilaton, are worth recalling. First, there is the well known
problem of dilaton--stabilization. Regardless of whether
SUSY--breaking utilizes the VEV of the dilaton--auxiliary component, 
$F_S$, or not,
one needs to avoid its generic weak--coupling runaway behavior
({\it i.e.} $S\rightarrow\infty$), and obtain instead a
stable minimum of its potential at a value of $S=S_0\sim10-20$,
rather than at infinity or 1 [for a discussion of this issue
and references to various attempts for its resolution in the
field--theory and string--theory/M--theory context, see ${\it e.g.}$
Ref. \cite{stabDVEV} and references therein].
Second, if SUSY--breaking is dominated by the VEV of $F_S$, it seems
that one would encounter the problem of color and
electric charge -- breakings (see {\it e.g.} last paper of Ref. \cite{DDSB}).

It seems to us, however, that in a mutually coupled
system such as ours, supersymmetry--breaking may well proceed
through multiple sources whose effects on
soft masses may in general be comparable.
In some cases, one of the sources may be viewed as primary
and the other(s) secondary, and the former may in fact induce
the latter.
We have in mind the
mutual couplings between (a) the dilaton superfield $S$ and the
non--perturbatively generated hidden sector $SU(5)_H$
gaugino and matter condensates (whose scale $\Lambda_5$ is
proportional to ${\rm e}^{(-S/2b_5)}$) on the one hand, and (b)
that between $S$ and the anomalous $U(1)_A$ gauge field
via the Green--Schwarz term that generates the Fayet--Iliopoulos
term $\xi$ on the other hand (see {\it e.g.} Ref. \cite{ua1}).
{\it Because of the mutual couplings,
these three components -- {\it i.e.} the dilaton,
the hidden sector condensates and the anomalous $U(1)$ --
can influence each other's role significantly
and thereby the nature of SUSY--breaking}.
The task at hand therefore is the minimization of the full
effective potential for this coupled system, which receives
contributions from (a) and (b), as well as possibly from additional
non--perturbative terms in the K\"ahler potential.
Such a minimization is to be carried out in the
presence of the SUSY--preserving VEVs of standard model
singlet fields $\{\Phi^i\}$ (see {\it e.g.} (\ref{patternsol2})),
which are induced because of the Fayet--Iliopoulos term,
and which generate the mass term $m$, by utilizing the hidden sector
condensates, as in section 4.

One particularly attractive possibility which we defer
for further study is this. The hidden sector condensates,
involving in general matter and gaugino pairs, in conjunction
with the SUSY--preserving VEVs of the $\{\Phi_i\}$--fields
generate the mass term $m$ as in section 4, which in turn
triggers SUSY--breaking through a family universal $U(1)_A$,
as in section 3. This could provide at least a major contribution to
squark--masses $\sim~(1/3-2){\rm TeV}$ (say),
which is approximately family universal.
Simultaneously, the coupling of the dilaton to the hidden
sector condensates as well as to the gauge field of $U(1)_A$,
together perhaps with non--perturbative terms in the K\"ahler
potential,
stabilizes the dilaton at a desired value $S_0$,
while inducing a VEV for the dilaton auxiliary field
$\langle F_{S_0}\rangle\ne0$.

Such a scenario, if it can be realized, would have the following
advantages: (i) The dilaton--induced SUSY breaking
($\langle F_S\rangle\ne0$) would not upset
the squark--degeneracy
that was obtained through the family--universal $U(1)_A$, because
the dilaton contributes universally to the
scalar masses (barring smaller loop--corrections).
(ii) Since dilaton SUSY--breaking
assigns comparable masses to squarks and gauginos
(unlike $U(1)_A$), following the relations $(\Delta
m_{\tilde g})_{F_S}=\sqrt{3}(\delta m_{\tilde q})_{F_S}=
\sqrt{3}(m_{3/2})_{F_S}$, however, it could provide the
leading contribution to gluino and wino masses:
$(\Delta m_{\tilde g})_{F_S}\approx\sqrt{3}
(1~{\rm to~few})(100){\rm GeV}$ (say) -- while providing
significant contributions to squark--masses\footnote{While $\langle F_S\rangle$
and $\langle D_A\rangle$--contributions to squark--masses
may be comparable, their relative proportion
will be constrained by the need to avoid color and charge--breaking
(see Ref. \cite{DDSB}). Quite clearly, dominant contribution
from $\langle F_S\rangle$ would be excluded on this ground.}.
This would remove the problem
of light gluino for $U(1)_A$ -- SUSY breaking.
(iii) SUSY breaking through {\it the combination of
a universal $U(1)_A$ and dilaton mechanisms}, as described above,
would of course avoid the danger of color and
charge--breaking, that confronts the scenario
of purely dilaton dominated SUSY breaking
(see the last paper in Ref. \cite{DDSB}).
In short, the combined SUSY--breaking
mechanism involving a family--universal $U(1)_A$ and the dilaton
has the advantage
that each component cures the vices of the other, without
upsetting any of its virtues.
The feasibility of this combined source of
SUSY--breaking, including
its effects on dilaton -- stabilization, is
clearly worth further study \cite{lalak}.

(3) Unlike the models of Refs. \cite{dvali} and \cite{mohapatra},
which introduce very few fields and just the single anomalous
$U(1)_A$ , but without any accompanying flavor--symmetries, string
solutions generically contain many fields (see {\it e.g.} tables 1--3
and 4--6) and typically several $U(1)$'s, some of which are
family dependent. In spite of this more elaborate (though fixed)
structures, it is interesting that minimization for the
string solutions considered here, led to a hierarchical
pattern of soft masses (see Eq. (\ref{sqm2}))
which is very similar to the cases
of Refs. \cite{dvali}--\cite{mohapatra}, barring of course
the distinction of three versus two--family degeneracy that
arises from the differences in $U(1)_A$ (see section 5).
In particular, it is remarkable that, for the string solutions
considered here,
the VEVs of $D_\alpha$'s associated with family--dependent
$U(1)_\alpha$'s turned out to vanish, owing to the requirement
of a relatively global minimum of the potential.
But for this feature, SUSY--breaking
through anomalous $U(1)$ in string models would not be viable.

(4) $\underline{\hbox{{\bf The Necessary Ingredients}:}}$
{}From the preceding discussion and those in sections 3 and 4,
it is clear that the following set of ingredients
are in fact needed in order that supersymmetry breaking through
anomalous $U(1)$ can be implemented consistently,
especially in the string--context: (i) family--universality
of the $U(1)_A$ and therefore the family permutation symmetry
of the flavor gauge symmetries
as discussed above; (ii) suitably suppressed effective mass
term $m$ of relevant fields carrying the anomalous
charge; (iii) positivity of the anomalous charges
of the chiral squark and slepton fields; and (iv) vanishing (or adequate
suppression) of the undesirable $D$--terms, associated
with family--dependent $U(1)$'s, because of minimization of the
potential. It seems truly remarkable that there do exist string
solutions, as discussed here, for which all four
ingredients are realized. If the anomalous $U(1)$ would turn out to
provide an important
source of SUSY--breaking, realization of these necessary
features, as well as meeting the non--trivial constraints from
issues such as proton longevity \cite{custodial,jp} and fermion masses
and mixings, together would clearly provide a very useful set
of criteria in severely limiting the desired class of
solutions from the vast set that is available.

(5) $\underline{\hbox{{\bf Gravitino Mass}:}}$
An important effective parameter of SUSY--breaking
is the mass of the gravitino.
With SUSY--breaking through only anomalous $U(1)$,
as described in section 3, the gravitino
would receive a mass $m_{3/2}\sim\langle F({\bar\Phi}_{45})\rangle/M_{Pl}
\approx m\sqrt{\epsilon}\approx(m_{\tilde q}/10)\sim(1~{\rm to~few})(100
{\rm GeV})$. With additional sources of SUSY--breaking, involving
for example the dilaton and possibly hidden sector
condensates, as motivated above, $m_{3/2}$ would get further
contributions. While the relative contributions
of these different sources of SUSY--breaking to $m_{3/2}$,
squark and gluino masses are not easy to ascertain at present,
we would still expect $m_{3/2}$ to lie in the 100GeV to a
few TeV--range.

\setcounter{footnote}{0}
\section{Summary}

In summary, an anomalous $U(1)$ gauge symmetry, together
with an effective mass--term for certain relevant fields,
offers a very simple mechanism to implement SUSY--breaking.
It is shown here that this mechanism can in fact be derived
consistently, leaving aside the question of dilaton--stabilization,
from an underlying string theory.
While string solutions invariably possess an
anomalous $U(1)$ symmetry, the requirement of three family
universality of squark masses and therefore of $U(1)_A$
is not easy to satisfy. 
We have shown that there
do exist certain three generation string solutions
for which supersymmetry breaking through an
anomalous $U(1)$ leads to both the desired three--family
squark--degeneracy
and the large hierarchy between the string and
the electroweak scales. More specifically we have noted that
these solutions, in contrast to most, possess {\it a cyclic permutation
symmetry} between the three families, which automatically yields
a set of non--anomalous but family--dependent flavor
gauge symmetries on the one hand, and a {\it family--universal
anomalous $U(1)$} gauge symmetry on the other hand. It
is the non--anomalous flavor symmetries, unavailable in GUTs,
which are ultimately responsible for hierarchical fermion masses
and CKM mixings as well as for the desired suppression
of both the rapid proton decay operators and of the effective
SUSY--breaking mass parameter $m$. The anomalous $U(1)_A$
is, however, family universal. In other words it is not
a horizontal symmetry, unlike the models of Refs. \cite{dvali} and
\cite{mohapatra}, and it is this feature that makes it suitable
for the purposes of SUSY--breaking without encountering color
and electric charge breaking.
We further note that family universality of the anomalous $U(1)$
has also been found to be desirable in recent attempts to
fit the fermion mass spectrum by the use of Abelian horizontal
symmetries \cite{ramond}.

We have remarked that the family permutation symmetry
of the solutions of interest \cite{eu,top}, is a joint
consequence of (a) the NAHE set of boundary condition
basis vectors which corresponds to a $Z_2\times Z_2$
orbifold compactification, and (b) the special choice
of additional boundary condition basis vectors, beyond the
NAHE set, which serve to reduce the number of generations
from 48 to 3. While suitable variations in (b) could still
allow the permutation symmetry to be retained, it is far from clear
whether the same can still be realized without the NAHE set.

As regards the issue of supersymmetry--breaking, we have
noted that the marriage of the two sources for such a breaking -- {\it i.e.}
through $U(1)_A$ and through the dilaton -- if it can be realized,
would be most attractive because it would combine the advantages of both,
while each would remove the disadvantage of the other.
Realization of this combined mechanism,
would thus be of major importance.

To conclude, if the $D$--term of the anomalous $U(1)$ makes a major
contribution to squark masses, it must be family--universal. 
In this case, if the NAHE set turns out to be a necessary
ingredient to obtain a family universal anomalous $U(1)$,
it would be an indication that the string vacuum
is in the vicinity of the $Z_2\times Z_2$ orbifold,
with the standard embedding of the gauge connection. Thus,
the question about the necessity of the NAHE set for
obtaining the family--permutation symmetry is
an interesting and important one, worth further study.


\bigskip
It is a pleasure to thank I. Antoniadis, G. Cleaver,
C. Kolda, J. March--Russell, A. Pomarol, P. Ramond and especially K.S. Babu,
G. Dvali  and H. Murayama for useful discussions.
We also wish to thank the hospitality of the CERN
theory division, where part of this work was done.
This work was supported in part by
DOE Grant No.\ DE-FG-0586ER40272 (AEF) and in part by
PHY--9119745 (JCP). The research of JCP was supported
in part by a Distinguished Faculty Research Fellowship
awarded by the University of Maryland.

$\underline{\hbox{{\bf Note added in proof}:}}$

After the submission of our paper, an interesting work has recently
appeared \cite{adm}, where the authors demonstrate the emergence
of the {\it combined} anomalous $U(1)$--dilaton
SUSY--breaking picture ({\it i.e.} $\langle D_A\rangle\ne0$
{\it and} $\langle F_S\rangle\ne0$), which was motivated
in sec. 6 of our paper, because of the mutual coupling
between the two systems and also on phenomenological grounds.
While the dilaton--contribution seems to dominate
over that of $\langle D_A\rangle$ in the cases
studied in \cite{adm}, as the authors note, the
relative contributions of $\langle D_A\rangle$ and
$\langle F_S\rangle$ to squark masses would of course depend
upon the manner of dilaton--stabilization.
Following remarks in sec. 6, a desirable solution
would seem to be one in which the $\langle D_A\rangle$
contribution is at least comparable to that of $\langle F_S\rangle$,
so that the problem of color--charge breaking is avoided.
This issue needs further study.

\bibliographystyle{unsrt}

\vfill
\eject

\textwidth=7.5in
\oddsidemargin=-18mm
\topmargin=-5mm
\renewcommand{\baselinestretch}{1.3}
\pagestyle{empty}
\begin{table}
\begin{eqnarray*}
\begin{tabular}{|c|c|c|rrrrrrrr|c|rr|}
\hline
  $F$ & SEC & $SU(3)\times SU(2)$&$Q_{C}$ & $Q_L$ & $Q_1$ &
   $Q_2$ & $Q_3$ & $Q_{4}$ & $Q_{5}$ & $Q_6$ &
   $SU(5)\times SU(3)$ & $Q_{7}$ & $Q_{8}$ \\
\hline
   $L_1$ & $b_1$      & $(1,2)$ & $-{3\over2}$ & $0$ &
   ${1\over2}$ & $0$ & $0$ & ${1\over 2}$ & $0$ & $0$ &
   $(1,1)$ & $0$ & $0$ \\
   $Q_1$ &            & $(3,2)$ & $ {1\over2}$ & $0$ &
   ${1\over2}$ & $0$ & $0$ & $-{1\over 2}$ & $0$ & $0$ &
   $(1,1)$ & $0$ & $0$ \\
   $d_1$ &            & $({\bar 3},1)$ & $-{1\over2}$ & $-1$ &
   ${1\over2}$ & $0$ & $0$ & $-{1\over2}$ & $0$ & $0$ &
   $(1,1)$ & $0$ & $0$ \\
   ${N}_1$ &            & $(1,1)$ & ${3\over2}$ & $-1$ &
   ${1\over 2}$ & $0$ & $0$ & $-{1\over 2}$ & $0$ & $0$ &
   $(1,1)$ & $0$ & $0$ \\
   $u_1$ &            & $({\bar 3},1)$ & $-{1\over2}$ & $1$ &
   ${1\over2}$ & $0$ & $0$ & ${1\over2}$ & $0$ & $0$ &
   $(1,1)$ & $0$ & $0$ \\
   ${e}_1$ &            & $(1,1)$ & ${3\over2}$ & $1$ &
   ${1\over 2}$ & $0$ & $0$ & ${1\over 2}$ & $0$ & $0$ &
   $(1,1)$ & $0$ & $0$ \\
\hline
   $L_2$ & $b_2$      & $(1,2)$ & $-{3\over2}$ & $0$ &
   $0$ & ${1\over2}$ & $0$ & $0$ & ${1\over 2}$ & $0$ &
   $(1,1)$ & $0$ & $0$ \\
   $Q_2$ &            & $(3,2)$ & $ {1\over2}$ & $0$ &
   $0$ & ${1\over2}$ & $0$ & $0$ & $-{1\over 2}$ & $0$ &
   $(1,1)$ & $0$ & $0$ \\
   $d_2$ &            & $({\bar 3},1)$ & $-{1\over2}$ & $-1$ &
   $0$ & ${1\over2}$ & $0$ & $0$ & $-{1\over2}$ & $0$ &
   $(1,1)$ & $0$ & $0$ \\
   ${N}_2$ &            & $(1,1)$ & ${3\over2}$ & $-1$ &
   $0$ & ${1\over 2}$ & $0$ & $0$ & $-{1\over 2}$ & $0$ &
   $(1,1)$ & $0$ & $0$ \\
   $u_2$ &            & $({\bar 3},1)$ & $-{1\over2}$ & $1$ &
   $0$ & ${1\over2}$ & $0$ & $0$ & ${1\over2}$ & $0$ &
   $(1,1)$ & $0$ & $0$ \\
   ${e}_2$ &            & $(1,1)$ & ${3\over2}$ & $1$ &
   $0$ & ${1\over 2}$ & $0$ & $0$ & ${1\over 2}$ & $0$ &
   $(1,1)$ & $0$ & $0$ \\
\hline
   $L_3$ & $b_3$      & $(1,2)$ & $-{3\over2}$ & $0$ &
   $0$ & $0$ & ${1\over2}$ & $0$ & $0$ & ${1\over 2}$ &
   $(1,1)$ & $0$ & $0$ \\
   $Q_3$ &            & $(3,2)$ & $ {1\over2}$ & $0$ &
   $0$ & $0$ & ${1\over2}$ & $0$ & $0$ & $-{1\over 2}$ &
   $(1,1)$ & $0$ & $0$ \\
   $d_3$ &            & $({\bar 3},1)$ & $-{1\over2}$ & $-1$ &
   $0$ & $0$ & ${1\over2}$ & $0$ & $0$ & $-{1\over2}$ &
   $(1,1)$ & $0$ & $0$ \\
   ${N}_3$ &            & $(1,1)$ & ${3\over2}$ & $-1$ &
   $0$ & $0$ & ${1\over 2}$ & $0$ & $0$ & $-{1\over 2}$ &
   $(1,1)$ & $0$ & $0$ \\
   $u_3$ &            & $({\bar 3},1)$ & $-{1\over2}$ & $1$ &
   $0$ & $0$ & ${1\over2}$ & $0$ & $0$ & ${1\over2}$ &
   $(1,1)$ & $0$ & $0$ \\
   ${e}_3$ &            & $(1,1)$ & ${3\over2}$ & $1$ &
   $0$ & $0$ & ${1\over 2}$ & $0$ & $0$ & ${1\over 2}$ &
   $(1,1)$ & $0$ & $0$ \\
\hline
   $h_1$ & ${\rm NS}$ & $(1,2)$ & $0$ & $-1$ &
   $1$ & $0$ & $0$ & $0$ & $0$ & $0$ &
   $(1,1)$ & $0$ & $0$ \\
   $h_2$ &                       & $(1,2)$ & $0$ & $-1$ &
   $0$ & $1$ & $0$ & $0$ & $0$ & $0$ &
   $(1,1)$ & $0$ & $0$ \\
   $h_3$ &                       & $(1,2)$ & $0$ & $-1$ &
   $0$ & $0$ & $1$ & $0$ & $0$ & $0$ &
   $(1,1)$ & $0$ & $0$ \\
   $\Phi_{12}$ &                 & $(1,1)$ & $0$ & $0$ &
   $1$ & $-1$ & $0$ & $0$ & $0$ & $0$ &
   $(1,1)$ & $0$ & $0$ \\
   $\Phi_{13}$ &                 & $(1,1)$ & $0$ & $0$ &
   $1$ & $0$ & $-1$ & $0$ & $0$ & $0$ &
   $(1,1)$ & $0$ & $0$ \\
   $\Phi_{23}$ &                 & $(1,1)$ & $0$ & $0$ &
   $0$ & $1$ & $-1$ & $0$ & $0$ & $0$ &
   $(1,1)$ & $0$ & $0$ \\
\hline
   $h_{45}$ & $b_1+b_2+$ & $(1,2)$ & $0$ & $-1$ &
   $-{1\over2}$ & $-{1\over2}$ & $0$ & $0$ & $0$ & $0$ &
   $(1,1)$ & $0$ & $0$ \\
   $D_{45}$ & $\alpha+\beta$ & $(3,1)$ & $-1$ & $0$ &
   $-{1\over2}$ & $-{1\over2}$ & $0$ & $0$ & $0$ & $0$ &
   $(1,1)$ & $0$ & $0$ \\
   $\Phi_{45}$ &                & $(1,1)$ & $0$ & $0$ &
   $-{1\over2}$ & $-{1\over2}$ & $-1$ & $0$ & $0$ & $0$ &
   $(1,1)$ & $0$ & $0$ \\
   $\Phi_{1}^\pm$ &              & $(1,1)$ & $0$ & $0$ &
   $-{1\over2}$ & ${1\over2}$ & $0$ & $\pm1$ & $0$ & $0$ &
   $(1,1)$ & $0$ & $0$ \\
   $\Phi_{2}^\pm$ &              & $(1,1)$ & $0$ & $0$ &
   $-{1\over2}$ & ${1\over2}$ & $0$ & $0$ & $\pm1$ & $0$ &
   $(1,1)$ & $0$ & $0$ \\
   $\Phi_{3}^\pm$ &              & $(1,1)$ & $0$ & $0$ &
   $-{1\over2}$ & ${1\over2}$ & $0$ & $0$ & $0$ & $\pm1$ &
   $(1,1)$ & $0$ & $0$ \\
\hline
\end{tabular}
\label{matter1}
\end{eqnarray*}
\caption{Massless states for solution I (Ref. \cite{eu})
which transform solely under the
observable gauge group.
$Q_C=3/2(B-L)$ and $Q_L=2T_{3_R}$.
In the NS and the $b_1+b_2+\alpha+\beta$
sectors contain also the conjugate states (${\bar h}_1$, etc.).
The NS sector contains additional three singlet states, $\xi_{1,2,3}$,
which are neutral under all the $U(1)$ symmetries.}
\end{table}
\vfill
\eject

\begin{table}
\begin{eqnarray*}
\begin{tabular}{|c|c|c|rrrrrrrr|c|rr|}
\hline
  $F$ & SEC & $SU(3)\times SU(2)$&$Q_{C}$ & $Q_L$ & $Q_1$ &
   $Q_2$ & $Q_3$ & $Q_{4}$ & $Q_{5}$ & $Q_6$ &
   $SU(5)\times SU(3)$ & $Q_{7}$ & $Q_{8}$ \\
\hline
   $V_1$ & $b_1+2\gamma+$ & $(1,1)$ & $0$ & $0$ & $0$ & ${1\over 2}$ &
   ${1\over2}$ & ${1\over 2}$ & $0$ & $0$ & $(1,3)$ & $-{1\over 2}$ &
   ${5\over2}$ \\
   ${\bar V}_1$ & $(I)$ & $(1,1)$ & $0$ & $0$ & $0$ & ${1\over 2}$ &
   ${1\over2}$ & ${1\over 2}$ & $0$ & $0$ & $(1,{\bar3})$ & ${1\over 2}$ &
   $-{5\over2}$ \\
   $T_1$ &                   & $(1,1)$ & $0$ & $0$ & $0$ & ${1\over2}$ &
   ${1\over2}$ & $-{1\over2}$ & $0$ & $0$ & $(5,1)$ & $-{1\over 2}$ &
   $-{3\over2}$ \\
   ${\bar T}_1$ &            & $(1,1)$ & $0$ & $0$ & $0$ & ${1\over 2}$ &
   ${1\over 2}$ & $-{1\over 2}$ & $0$ & $0$ & $({\bar 5},1)$ & ${1\over 2}$ &
   ${3\over2}$ \\
\hline
   $V_2$ & $b_2+2\gamma+$ & $(1,1)$ & $0$ & $0$ & ${1\over 2}$ &
    $0$ & ${1\over 2}$ & $0$ & ${1\over 2}$ & $0$ & $(1,3)$ & $-{1\over 2}$ &
   ${5\over2}$ \\
   ${\bar V}_2$ & $(I)$      & $(1,1)$ & $0$ & $0$ & ${1\over 2}$ &
    $0$ & ${1\over 2}$ & $0$ & ${1\over 2}$ & $0$ & $(1,{\bar3})$ &
    ${1\over 2}$ &
    $-{5\over2}$ \\
   $T_2$ &  		     & $(1,1)$ & $0$ & $0$ & ${1\over 2}$ &
    $0$ & ${1\over 2}$ & $0$ & $-{1\over2}$ & $0$ & $(5,1)$ & $-{1\over 2}$ &
   $-{3\over2}$ \\
   ${\bar T}_2$ &  	     & $(1,1)$ & $0$ & $0$ & ${1\over 2}$ &
    $0$ & ${1\over 2}$ & $0$ & $-{1\over 2}$ & $0$ & $({\bar5},1)$ &
    ${1\over 2}$ &
    ${3\over2}$ \\
\hline
   $V_3$ & $b_3+\gamma+$ & $(1,1)$ & $0$ & $0$ & ${1\over 2}$ &
   ${1\over2}$ & $0$ & $0$ & $0$ & ${1\over2}$ & $(1,3)$ & $-{1\over 2}$ &
   ${5\over2}$ \\
   ${\bar V}_3$ & $(I)$ & $(1,1)$ & $0$ & $0$ & ${1\over2}$ &
   ${1\over2}$ & $0$ & $0$ & $0$ & ${1\over2}$ & $(1,{\bar3})$ & ${1\over 2}$ &
   $-{5\over2}$ \\
   $T_3$ &  		     & $(1,1)$ & $0$ & $0$ & ${1\over2}$ &
   ${1\over2}$ & $0$ & $0$ & $0$ & $-{1\over 2}$ & $(5,1)$ & $-{1\over 2}$ &
   $-{3\over2}$ \\
   ${\bar T}_3$ &  	     & $(1,1)$ & $0$ & $0$ & ${1\over 2}$ &
   ${1\over2}$ & $0$ & $0$ & $0$ & $-{1\over 2}$ & $({\bar5},1)$ &
   ${1\over 2}$ &
   ${3\over2}$  \\
\hline
   $H_{1}$ & $b_1+b_2+$     & $(1,1)$ & ${3\over4}$ & ${1\over2}$ &
   $-{1\over4}$ & $-{1\over 4}$ & ${1\over 4}$ &
   $-{1\over2}$ & $-{1\over 2}$ & $0$ &
   (1,3) & ${1\over4}$ & $-{5\over4}$   \\
   $H_{2}$ & $\alpha+\beta$ & (1,1) & $-{3\over4}$  & $-{1\over2}$ &
   ${1\over4}$ & ${1\over 4}$ & $-{1\over4}$ &
   $-{1\over2}$ & $-{1\over 2}$ & $0$ &
   (1,$\bar 3$) & $-{1\over 4}$ & ${5\over 4}$  \\
   $H_{3}$ &  $\pm\gamma+(I)$ & (1,1) & ${3\over4}$ & ${1\over2}$ &
   $-{1\over4}$ & $-{1\over 4}$ & ${1\over 4}$ &
   $-{1\over2}$ & $-{1\over 2}$ & $0$ &
   (1,1) & $-{3\over 4}$ & ${15\over 4}$  \\
   $H_{4}$ &                  & (1,1) & $-{3\over4}$ & $-{1\over2}$ &
   ${1\over4}$ & ${1\over 4}$ & $-{1\over4}$ &
   $-{1\over2}$ & $-{1\over 2}$ & $0$ &
   (1,1) & $3\over4$ & $-{15\over 4}$  \\
\hline
   $H_{5}$ & $b_1+b_3+$     & $(1,1)$ & ${3\over4}$ & ${1\over2}$ &
   $-{1\over4}$ & ${1\over 4}$ & $-{1\over 4}$ &
   $-{1\over2}$ & $0$ & $-{1\over 2}$ &
   (1,3) & ${1\over4}$ & $-{5\over4}$   \\
   $H_{6}$ & $\alpha+\beta$ & (1,1) & $-{3\over4}$  & $-{1\over2}$ &
   ${1\over4}$ & $-{1\over 4}$ & ${1\over4}$ &
   $-{1\over2}$ & $0$ & $-{1\over 2}$ &
   (1,$\bar 3$) & $-{1\over 4}$ & ${5\over 4}$  \\
   $H_{7}$ &  $\pm\gamma+(I)$ & (1,1) & ${3\over4}$ & ${1\over2}$ &
   $-{1\over4}$ & ${1\over 4}$ & $-{1\over 4}$ &
   $-{1\over2}$ & $0$ & $-{1\over 2}$ &
   (1,1) & $-{3\over 4}$ & ${15\over 4}$  \\
   $H_{8}$ &                  & (1,1) & $-{3\over4}$ & $-{1\over2}$ &
   ${1\over4}$ & $-{1\over 4}$ & ${1\over4}$ &
   $-{1\over2}$ & $0$ & $-{1\over 2}$ &
   (1,1) & $3\over4$ & $-{15\over 4}$  \\
\hline
   $H_{9}$ & $b_2+b_3+$     & $(1,1)$ & ${3\over4}$ & ${1\over2}$ &
   ${1\over4}$ & $-{1\over 4}$ & $-{1\over 4}$ &
   $0$ & $-{1\over2}$ & $-{1\over 2}$ &
   (1,3) & ${1\over4}$ & $-{5\over4}$   \\
   $H_{10}$ & $\alpha+\beta$ & (1,1) & $-{3\over4}$  & $-{1\over2}$ &
   $-{1\over4}$ & $-{1\over 4}$ & ${1\over4}$ &
   $0$ & $-{1\over2}$ & $-{1\over 2}$ &
   (1,$\bar 3$) & $-{1\over 4}$ & ${5\over 4}$  \\
   $H_{11}$ &  $\pm\gamma+(I)$ & (1,1) & ${3\over4}$ & ${1\over2}$ &
   ${1\over4}$ & $-{1\over 4}$ & $-{1\over 4}$ &
   $0$ & $-{1\over2}$ & $-{1\over 2}$ &
   (1,1) & $-{3\over 4}$ & ${15\over 4}$  \\
   $H_{12}$ &                  & (1,1) & $-{3\over4}$ & $-{1\over2}$ &
   $-{1\over4}$ & $-{1\over 4}$ & ${1\over4}$ &
   $0$ & $-{1\over2}$ & $-{1\over 2}$ &
   (1,1) & $3\over4$ & $-{15\over 4}$  \\
\hline
   $H_{13}$ & $b_1+b_3+$ & $(1,1)$ & $-{3\over4}$ & ${1\over2}$ &
   $-{1\over4}$ & ${1\over 4}$ & $-{1\over 4}$ & 0 & 0 & 0 &
   (1,3) & ${3\over4}$ & $5\over4$   \\
   $H_{14}$ &  $\alpha\pm\gamma+$  & (1,1) & ${3\over4}$  & $-{1\over2}$ &
   ${1\over4}$ & $-{1\over 4}$ & $1\over 4$ & 0 & 0 & 0 &
   (1,$\bar 3$) & $-{3\over 4}$ & $-{5\over 4}$  \\
   $H_{15}$ &     $(I)$       & (1,2) & $-{3\over4}$ & $-{1\over2}$ &
   $-{1\over4}$ & ${1\over 4}$ & $-{1\over 4}$ & 0 & 0 & 0 &
   (1,1) & $-{1\over 4}$ & $-{15\over 4}$  \\
   $H_{16}$ &               & (1,2) & ${3\over4}$ & $1\over2$ &
   ${1\over4}$ & $-{1\over 4}$ & $1\over 4$ & 0 & 0 & 0 &
   (1,1) & $1\over4$ & ${15\over 4}$  \\
   $H_{17}$ &               &  (1,1) & $-{3\over4}$ & $1\over2$ &
   $-{1\over 4}$ & $-{3\over4}$ & $-{1\over 4}$ & 0 & 0 & 0 &
   (1,1) & $-{1\over 4}$ & $-{15\over 4}$  \\
   $H_{18}$ &               & (1,1) & $3\over4$ & $-{1\over2}$ &
   ${1\over 4}$ & $3\over4$ & $1\over 4$ & 0 & 0 & 0 &
   (1,1) & ${1\over 4}$ & ${15\over 4}$  \\
\hline
\end{tabular}
\label{matter2}
\end{eqnarray*}
\caption{Massless states for solution I (\cite{eu}).}
\end{table}

\vfill
\eject

\begin{table}
\begin{eqnarray*}
\begin{tabular}{|c|c|c|rrrrrrrr|c|rr|}
\hline
  $F$ & SEC & $SU(3)\times SU(2)$&$Q_{C}$ & $Q_L$ & $Q_1$ &
   $Q_2$ & $Q_3$ & $Q_{4}$ & $Q_{5}$ & $Q_6$ &
   $SU(5)\times SU(3)$ & $Q_{7}$ & $Q_{8}$ \\
\hline
   $H_{19}$ & $b_2+b_3+$ & (1,1) & $-{3\over4}$ & $1\over2$ &
   ${1\over 4}$ & $-{1\over4}$ & $-{1\over 4}$ & 0 & 0 & 0 &
   (5,1)&  $-{1\over 4}$ & ${9\over 4}$ \\
   $H_{20}$ & $\alpha\pm\gamma+$  & (1,1) & $3\over4$ & $-{1\over2}$ &
   $-{1\over 4}$ & $1\over4$ & $1\over 4$ & 0 & 0 & 0 &
   ($\bar 5$,1) & $1\over 4$ & $-{9\over 4}$  \\
   $H_{21}$ &  $(I)$         & (3,1) & $1\over4$ & $1\over2$ &
   ${1\over 4}$ & $-{1\over 4}$ & $-{1\over4}$ & 0 & 0 & 0 &
   (1,1) & $-{1\over 4}$ & $-{15\over 4}$  \\
   $H_{22}$ &                & ($\bar 3$,1) & $-{1\over4}$ & $-{1\over2}$
   & $-{1\over 4}$ & ${1\over 4}$ & $1\over4$ & 0 & 0 & 0 &
   (1,1) & ${1\over 4}$ & ${15\over 4}$  \\
   $H_{23}$ &                & (1,1) & $-{3\over4}$ & $1\over2$ &
   ${1\over 4}$ & $-{1\over 4}$ & $3\over4$ & 0 & 0 & 0 &
   (1,1) & ${1\over 4}$ & ${15\over 4}$  \\
   $H_{24}$ &                & (1,1) & $3\over4$ & $-{1\over2}$ &
   $-{1\over4}$ & ${1\over 4}$ & $-{3\over4}$ & 0 & 0 & 0 &
   (1,1) & $-{1\over 4}$ & $-{15\over 4}$ \\
   $H_{25}$ &                & (1,1) & $-{3\over4}$ & $ 1\over2$ &
   $1\over4$ & $3\over4$ & $-{1\over4}$ & 0 & 0 & 0 &
   (1,1) & $-{1\over4}$ & $-{15\over4}$ \\
   $H_{26}$ &                & (1,1) & $3\over4$ &  $-{1\over2}$ &
   $-{1\over4}$ &  $-{3\over4}$ & $1\over4$ & 0 & 0 & 0  &
   (1,1) & $1\over4$ & $15\over4$  \\
\hline
   $H_{27}$ & $b_1+b_2+$     & $(1,1)$ & $-{3\over4}$ & $-{1\over2}$ &
   $-{1\over4}$ & $-{1\over 4}$ & $-{1\over 4}$ &
   $-{1\over2}$ & ${1\over 2}$ & ${1\over 2}$ &
   (1,3) & ${1\over4}$ & $-{5\over4}$   \\
   $H_{28}$ & $b_3+\alpha+$ & (1,1) & ${3\over4}$  & ${1\over2}$ &
   ${1\over4}$ & ${1\over 4}$ & ${1\over4}$ &
   ${1\over2}$ & $-{1\over 2}$ & $-{1\over 2}$ &
   (1,$\bar 3$) & $-{1\over 4}$ & ${5\over 4}$  \\
   $H_{29}$ &  $\beta\pm\gamma+$ & (1,1) & $-{3\over4}$ & $-{1\over2}$ &
   $-{1\over4}$ & $-{1\over 4}$ & $-{1\over 4}$ &
   ${1\over2}$ & $-{1\over 2}$ & ${1\over 2}$ &
   (1,1) & $-{3\over 4}$ & ${15\over 4}$  \\
   $H_{30}$ & $(I)$               & (1,1) & ${3\over4}$ & ${1\over2}$ &
   ${1\over4}$ & ${1\over 4}$ & ${1\over4}$ &
   $-{1\over2}$ & ${1\over 2}$ & $-{1\over 2}$ &
   (1,1) & $3\over4$ & $-{15\over 4}$  \\
\hline
\end{tabular}
\label{matter3}
\end{eqnarray*}
\caption{Massless states for solution I (\cite{eu}).}
\end{table}

\begin{table}
\begin{eqnarray*}
\begin{tabular}{|c|c|c|rrrrrrr|c|rr|}
\hline
   $F$ & SEC & $SU(4)_C\times SU(2)_L$&$Q_{C'}$ & $Q_L$ & $Q_1$ &
   $Q_2$ & $Q_3$ & $Q_{4'}$ & $Q_{5'}$ & $SU(5)_H\times SU(3)_H$ &
   $Q_{6'}$ & $Q_{8''}$ \\
\hline
   $L_1$ & $b_1 \oplus$ & $(1,2)$&$-{3\over 2}$ & $0$ & ${1\over 2}$ &
   $0$ & $0$ & $-{1\over 2}$ & $-{1\over 2}$ & $(1,1)$ & $-{8\over 3}$ &
   $0$ \\
   $Q_1$ & $1+\alpha+2\gamma$&$(4,2)$&${1\over 2}$&$0$&${1\over 2}$ &
   $0$ & $0$ & $-{1\over 2}$ & $-{1\over 2}$ & $(1,1)$ & $-{2\over 3}$ &
   $0$ \\
   $d_1$ &  & $(\overline 4,1)$&$-{1\over 2}$ & $1$ & ${1\over 2}$ &
   $0$ & $0$ & ${1\over 2}$ & ${1\over 2}$ & $(1,1)$ & ${2\over 3}$ &
   $0$ \\
   $N_1$ &  & $(1,1)$&${3\over 2}$ & $-1$ & ${1\over 2}$ &
   $0$ & $0$ & ${1\over 2}$ & ${1\over 2}$ & $(1,1)$ & ${8\over 3}$ &
   $0$ \\
   $e_1$ &  & $(1,1)$&${3\over 2}$ & $1$ & ${1\over 2}$ &
   $0$ & $0$ & ${1\over 2}$ & ${1\over 2}$ & $(1,1)$ & ${8\over 3}$ &
   $0$ \\
   $u_1$ &  & $(\overline 4,1)$&$-{1\over 2}$ & $-1$ & ${1\over 2}$ &
   $0$ & $0$ & ${1\over 2}$ & ${1\over 2}$ & $(1,1)$ & ${2\over 3}$ &
   $0$ \\
\hline
   $L_2$ & $b_2 \oplus$ & $(1,2)$&$-{3\over 2}$ & $0$&0 & ${1\over 2}$ &
    $0$ & ${1\over 2}$ & $-{1\over 2}$ & $(1,1)$ & $-{8\over 3}$ &
   $0$ \\
   $Q_2$ & $1+\alpha+2\gamma$&$(4,2)$&${1\over 2}$&$0$&0&${1\over 2}$ &
    $0$ & ${1\over 2}$ & $-{1\over 2}$ & $(1,1)$ & $-{2\over 3}$ &
   $0$ \\
   $d_2$ &  & $(\overline 4,1)$&$-{1\over 2}$ & $1$&0 & ${1\over 2}$ &
    $0$ & $-{1\over 2}$ & ${1\over 2}$ & $(1,1)$ & ${2\over 3}$ &
   $0$ \\
   $N_2$ &  & $(1,1)$&${3\over 2}$ & $-1$ & 0 & ${1\over 2}$ &
    $0$ & $-{1\over 2}$ & ${1\over 2}$ & $(1,1)$ & ${8\over 3}$ &
   $0$ \\
   $e_2$ &  & $(1,1)$&${3\over 2}$ & $1$&0 & ${1\over 2}$ &
    $0$ & $-{1\over 2}$ & ${1\over 2}$ & $(1,1)$ & ${8\over 3}$ &
   $0$ \\
   $u_2$ &  & $(\overline 4,1)$&$-{1\over 2}$ & $-1$&0 & ${1\over 2}$ &
    $0$ & $-{1\over 2}$ & ${1\over 2}$ & $(1,1)$ & ${2\over 3}$ &
   $0$ \\
\hline
   $L_3$ & $b_3 \oplus$ & $(1,2)$&$-{3\over 2}$ & $0$&0&0 & ${1\over 2}$ &
   $0$ & ${1}$ & $(1,1)$ & $-{8\over 3}$ &
   $0$ \\
   $Q_3$ & $1+\alpha+2\gamma$&$(4,2)$&${1\over 2}$&$0$&0&0&${1\over 2}$ &
    $0$ & ${1}$ & $(1,1)$ & $-{2\over 3}$ &
   $0$ \\
   $d_3$ &  & $(\overline 4,1)$&$-{1\over 2}$ & $1$&0&0 & ${1\over 2}$ &
    $0$ & $-{1}$ & $(1,1)$ & ${2\over 3}$ &
   $0$ \\
   $N_3$ &  & $(1,1)$&${3\over 2}$ & $-1$&0&0 & ${1\over 2}$ &
    $0$ & $-{1}$ & $(1,1)$ & ${8\over 3}$ &
   $0$ \\
   $e_3$ &  & $(1,1)$&${3\over 2}$ & $1$&0&0 & ${1\over 2}$ &
    $0$ & $-{1}$ & $(1,1)$ & ${8\over 3}$ &
   $0$ \\
   $u_3$ &  & $(\overline 4,1)$&$-{1\over 2}$ & $-1$&0&0 & ${1\over 2}$ &
    $0$ & $-{1}$ & $(1,1)$ & ${2\over 3}$ &
   $0$ \\
\hline
   $h_1$ & ${\rm NS}$ & $(1,2)$ & $0$ & $-1$ & 1 & 0 & $0$ &
   $0$ & $0$ & $(1,1)$ & $0$  & $0$ \\
   $h_2$ &            & $(1,2)$ & $0$ & $-1$ & 0 & 1 & $0$ &
   $0$ & $0$ & $(1,1)$ & $0$  & $0$ \\
   $h_3$ &            & $(1,2)$ & $0$ & $-1$ & 0 & 0 & $1$ &
   $0$ & $0$ & $(1,1)$ & $0$ & $0$ \\
   $\Phi_{12}$ &      & $(1,1)$ & $0$ & $ 0$ & 1 & -1 & $0$ &
   $0$ & $0$ & $(1,1)$ & $0$ & $0$ \\
   $\Phi_{13}$ &       & $(1,1)$ & $0$ & $0$ & 1 & 0 & $-1$ &
   $0$ & $0$ & $(1,1)$ & $0$ & $0$ \\
   $\Phi_{23}$ &       & $(1,1)$ & $0$ & $0$ & 0 & 1 & $-1$ &
   $0$ & $0$ & $(1,1)$ & $0$ & $0$ \\
\hline
   $h_{45}$ & $b_1+b_2+$   & $(1,2)$ & $0$ & $-1$ &
   ${1\over2}$ & ${1\over2}$  & $0$ &
   $0$ & $0$ & $(1,1)$ & $0$  & $0$ \\
   $h_{45}^\prime$ &  $\alpha+\beta$ & $(1,2)$ & $0$ & $-1$ &
   $-{1\over 2}$ & $-{1\over 2}$ & $0$ &
   $0$ & $0$ & $(1,1)$ & $0$  & $0$ \\
   $\Phi_{45}$ &            & $(1,1)$ & $0$ & $0$ &
   $-{1\over2}$ & $-{1\over2}$ & $-1$ &
   $0$ & $0$ & $(1,1)$ & ${2\over 3}$ & $0$ \\
   $\Phi_{45}^\prime$ &      & $(1,1)$ & $0$ & $ 0$ &
   $-{1\over2}$ & $-{1\over2}$ & $1$ &
   $0$ & $0$ & $(1,1)$ & $0$ & $0$ \\
   $\Phi{1,2}$ &       & $(1,1)$ & $0$ & $0$ &
   $-{1\over2}$ & ${1\over2}$ & $0$ &
   $0$ & $0$ & $(1,1)$ & $0$ & $0$ \\

\hline
\end{tabular}
\label{sol2matter1}
\end{eqnarray*}
\caption{Massless states for solution II (ref. \cite{top}).}
\end{table}

\begin{table}
\begin{eqnarray*}
\begin{tabular}{|c|c|c|rrrrrrr|c|rr|}
\hline
   $F$ & SEC & $SU(4)_C\times SU(2)_L$&$Q_{C'}$ & $Q_L$ & $Q_1$ &
   $Q_2$ & $Q_3$ & $Q_{4'}$ & $Q_{5'}$ & $SU(5)_H\times SU(3)_H$ &
   $Q_{6'}$ & $Q_{8''}$ \\
\hline
    $V_1$ & $b_1+2\gamma$ & $(1,1)$&$-{1\over 2}$ & $0$ & $0$ &
   ${1\over 2}$ & ${1\over 2}$ & ${1\over 2}$ & ${1\over 2}$ &
   $(1,3)$ & ${8\over 3}$ & ${5\over 2}$ \\
    $\overline V_1$ & & $(1,1)$&${1\over 2}$ & $0$ & $0$ &
   ${1\over 2}$ & ${1\over 2}$ & $-{1\over 2}$ & $-{1\over 2}$ &
   $(1,{\bar3})$ & $-{8\over 3}$ & $-{5\over 2}$ \\
    $T_1$ &  & $(1,1)$&${1\over 2}$ & $0$ & $0$ &
   ${1\over 2}$ & ${1\over 2}$ & $-{1\over 2}$ & $-{1\over 2}$ &
   $({5},1)$ & $-{8\over 3}$ & ${3\over 2}$ \\
    $\overline T_1$ &  & $(1,1)$&$-{1\over 2}$ & $0$ & $0$ &
   ${1\over 2}$ & ${1\over 2}$ & ${1\over 2}$ & ${1\over 2}$ &
   $({\bar5,}1)$ & ${8\over 3}$ & $-{3\over 2}$ \\
\hline
    $V_2$ & $b_2+2\gamma$ & $(1,1)$&$-{1\over 2}$ & $0$ &
   ${1\over 2}$&0 & ${1\over 2}$ & $-{1\over 2}$ & ${1\over 2}$ &
   $(1,3)$ & ${8\over 3}$ & ${5\over 2}$ \\
    $\overline V_2$ & & $(1,1)$&${1\over 2}$ & $0$ &
   ${1\over 2}$&0 & ${1\over 2}$ & ${1\over 2}$ & $-{1\over 2}$ &
   $(1,{\bar3})$ & $-{8\over 3}$ & $-{5\over 2}$ \\
    $T_2$ &  & $(1,1)$&${1\over 2}$ & $0$ &
   ${1\over 2}$&0 & ${1\over 2}$ & ${1\over 2}$ & $-{1\over 2}$ &
   $(5,1)$ & $-{8\over 3}$ & ${3\over 2}$ \\
    $\overline T_2$ &  & $(1,1)$&$-{1\over 2}$ & $0$ &
   ${1\over 2}$&0 & ${1\over 2}$ & $-{1\over 2}$ & ${1\over 2}$ &
   $({\bar5},1)$ & ${8\over 3}$ & $-{3\over 2}$ \\
\hline
    $V_3$ & $b_3+2\gamma$ & $(1,1)$&$-{1\over 2}$ & $0$ &
   ${1\over 2}$ & ${1\over 2}$&0 & $0$ & $-1$ &
   $(1,3)$ & ${8\over 3}$ & ${5\over 2}$ \\
    $\overline V_3$ & & $(1,1)$&${1\over 2}$ & $0$ &
   ${1\over 2}$ & ${1\over 2}$&0 & $0$ & $1$ &
   $(1,{\bar3})$ & $-{8\over 3}$ & $-{5\over 2}$ \\
    $T_3$ &  & $(1,1)$&${1\over 2}$ & $0$ &
   ${1\over 2}$ & ${1\over 2}$&0 & $0$ & $1$ &
   $(5,1)$ & $-{8\over 3}$ & ${3\over 2}$ \\
    $\overline T_3$ &  & $(1,1)$&$-{1\over 2}$ & $0$ &
   ${1\over 2}$ & ${1\over 2}$&0 & $0$ & $-1$ &
   $({\bar5},1)$ & ${8\over 3}$ & $-{3\over 2}$ \\
\hline
   $l_1$ & $b_2+b_3+$ & $(1,2)$&$-{3\over 4}$ &
    $-{1\over 2}$ & ${1\over 4}$ & $-{1\over 4}$ & $-{1\over 4}$ &
    $0$ & $0$ & $(1,1)$ & $0$ & $-{15\over 4}$ \\
   $\overline l_1$ &$\beta+\gamma+\xi$  & $(1,\overline 2)$&${3\over 4}$ &
    ${1\over 2}$ & $-{1\over 4}$ & ${1\over 4}$ & ${1\over 4}$ &
    $0$ & $0$ & $(1,1)$ & $0$ & ${15\over 4}$ \\
   $S_1$ &  & $(1,1)$&${3\over 4}$ &
    $-{1\over 2}$ & $-{3\over 4}$ & $-{1\over 4}$ & $-{1\over 4}$ &
    $0$ & $0$ & $(1,1)$ & $0$ & $-{15\over 4}$ \\
   $\overline S_1$ &  & $(1,1)$&$-{3\over 4}$ &
    ${1\over 2}$ & ${3\over 4}$ & ${1\over 4}$ & ${1\over 4}$ &
    $0$ & $0$ & $(1,1)$ & $0$ & ${15\over 4}$ \\
   $S_2$ &  & $(1,1)$&${3\over 4}$ &
    $-{1\over 2}$ & $-{1\over 4}$ & $-{3\over 4}$ & ${1\over 4}$ &
    $0$ & $0$ & $(1,1)$ & $0$ & $-{15\over 4}$ \\
   $\overline S_2$ &  & $(1,1)$&$-{3\over 4}$ &
    ${1\over 2}$ & ${1\over 4}$ & ${3\over 4}$ & $-{1\over 4}$ &
    $0$ & $0$ & $(1,1)$ & $0$ & ${15\over 4}$ \\
   $S_3$ &  & $(1,1)$&${3\over 4}$ &
    $-{1\over 2}$ & $-{1\over 4}$ & ${1\over 4}$ & $-{3\over 4}$ &
    $0$ & $0$ & $(1,1)$ & $0$ & $-{15\over 4}$ \\
   $\overline S_3$ &  & $(1,1)$&$-{3\over 4}$ &
    ${1\over 2}$ & ${1\over 4}$ & $-{1\over 4}$ & ${3\over 4}$ &
    $0$ & $0$ & $(1,1)$ & $0$ & ${15\over 4}$ \\
   $H_1$ &  & $(1,1)$&$-{3\over 4}$ &
    ${1\over 2}$ & ${1\over 4}$ & $-{1\over 4}$ & $-{1\over 4}$ &
    $0$ & $0$ & $(5,1)$ & $0$ & ${9\over 4}$ \\
   $\overline H_1$ &  & $(1,1)$&${3\over 4}$ &
    $-{1\over 2}$ & $-{1\over 4}$ & ${1\over 4}$ & ${1\over 4}$ &
    $0$ & $0$ & $({\bar5},1)$ & $0$ & $-{9\over 4}$ \\
\hline
   $l_2$ & $b_1+b_3+$ & $(1,2)$&$-{3\over 4}$ &
    $-{1\over 2}$ & ${1\over 4}$ & $-{1\over 4}$ & $-{1\over 4}$ &
    $0$ & $0$ & $(1,1)$ & $0$ & $-{15\over 4}$ \\
   $\overline l_2$ & $\alpha+\gamma+\xi$  & $(1,\overline 2)$&${3\over 4}$ &
    ${1\over 2}$ & $-{1\over 4}$ & ${1\over 4}$ & ${1\over 4}$ &
    $0$ & $0$ & $(1,1)$ & $0$ & ${15\over 4}$ \\
   $S_4$ &  & $(1,1)$&${3\over 4}$ &
    $-{1\over 2}$ & $-{3\over 4}$ & $-{1\over 4}$ & ${1\over 4}$ &
    $0$ & $0$ & $(1,1)$ & $0$ & $-{15\over 4}$ \\
   $\overline S_4$ &  & $(1,1)$&$-{3\over 4}$ &
    ${1\over 2}$ & ${3\over 4}$ & ${1\over 4}$ & $-{1\over 4}$ &
    $0$ & $0$ & $(1,1)$ & $0$ & ${15\over 4}$ \\
   $S_5$ &  & $(1,1)$&${3\over 4}$ &
    $-{1\over 2}$ & $-{1\over 4}$ & $-{3\over 4}$ & $-{1\over 4}$ &
    $0$ & $0$ & $(1,1)$ & $0$ & $-{15\over 4}$ \\
   $\overline S_5$ &  & $(1,1)$&$-{3\over 4}$ &
    ${1\over 2}$ & ${1\over 4}$ & ${3\over 4}$ & ${1\over 4}$ &
    $0$ & $0$ & $(1,1)$ & $0$ & ${15\over 4}$ \\
   $S_6$ &  & $(1,1)$&${3\over 4}$ &
    $-{1\over 2}$ & ${1\over 4}$ & $-{1\over 4}$ & $-{3\over 4}$ &
    $0$ & $0$ & $(1,1)$ & $0$ & $-{15\over 4}$ \\
   $\overline S_6$ &  & $(1,1)$&$-{3\over 4}$ &
    ${1\over 2}$ & $-{1\over 4}$ & ${1\over 4}$ & ${3\over 4}$ &
    $0$ & $0$ & $(1,1)$ & $0$ & ${15\over 4}$ \\
   $H_2$ &  & $(1,1)$&$-{3\over 4}$ &
    ${1\over 2}$ & $-{1\over 4}$ & ${1\over 4}$ & $-{1\over 4}$ &
    $0$ & $0$ & $(5,1)$ & $0$ & ${9\over 4}$ \\
   $\overline H_2$ &  & $(1,1)$&${3\over 4}$ &
    $-{1\over 2}$ & ${1\over 4}$ & $-{1\over 4}$ & ${1\over 4}$ &
    $0$ & $0$ & $({\bar5},1)$ & $0$ & $-{9\over 4}$ \\
\hline
\end{tabular}
\label{sol2matter2}
\end{eqnarray*}
\caption{Massless states for solution II (ref. \cite{top}).}
\end{table}

\begin{table}
\begin{eqnarray*}
\begin{tabular}{|c|c|c|rrrrrrr|c|rr|}
\hline
   $F$ & SEC & $SU(4)_C\times SU(2)_L$&$Q_{C'}$ & $Q_L$ & $Q_1$ &
   $Q_2$ & $Q_3$ & $Q_{4'}$ & $Q_{5'}$ & $SU(5)_H\times SU(3)_H$ &
   $Q_{6'}$ & $Q_{8''}$ \\
\hline
   $l_4$ & $1+b_1+$ & $(1,2)$&$-1$ &
    $0$ & $-{1\over 2}$ & 0  & 0 &
    $-{1\over 2}$ & $-{1\over 2}$ & $(1,1)$ & ${16\over 3}$ & $0$ \\
   $S_7$ &$\alpha+2\gamma$  & $(1,1)$&$1$ &
    $1$ & $-{1\over 2}$ & $0$ & $0$ &
    ${1\over 2}$ & ${1\over 2}$ & $(1,1)$ & $-{16\over 3}$ & $0$ \\
   $\overline S_7$ &  & $(1,1)$&$1$ &
    $-1$ & $-{1\over 2}$ & $0$ & $0$ &
    ${1\over 2}$ & ${1\over 2}$ & $(1,1)$ & $-{16\over 3}$ & $0$ \\
\hline
   $l_5$ & $1+b_2+$ & $(1,2)$&$-1$ &
    $0$ & 0 & $-{1\over 2}$  & 0 &
    ${1\over 2}$ & $-{1\over 2}$ & $(1,1)$ & ${16\over 3}$ & $0$ \\
   $S_8$ &$\alpha+2\gamma$  & $(1,1)$&$1$ &
    $1$ & 0 & $-{1\over 2}$ & $0$ &
    $-{1\over 2}$ & ${1\over 2}$ & $(1,1)$ & $-{16\over 3}$ & $0$ \\
   $\overline S_8$ &  & $(1,1)$&$1$ &
    $-1$ & 0 & $-{1\over 2}$ & $0$ &
    $-{1\over 2}$ & ${1\over 2}$ & $(1,1)$ & $-{16\over 3}$ & $0$ \\
\hline
   $l_6$ & $1+b_3+$ & $(1,2)$&$-1$ &
    $0$ & $0$ & $0$ & $-{1\over 2}$ &
    $0$ & $1$ & $(1,1)$ & ${16\over 3}$ & $0$ \\
   $S_9$ &$\alpha+2\gamma$  & $(1,1)$&$1$ &
    $1$ & $0$ & $0$ & $-{1\over 2}$ &
    $0$ & $-1$ & $(1,1)$ & $-{16\over 3}$ & $0$ \\
   $\overline S_9$ &  & $(1,1)$&$1$ &
    $-1$ & $0$ & $0$ & $-{1\over 2}$ &
    $0$ & $-1$ & $(1,1)$ & $-{16\over 3}$ & $0$ \\
\hline
   $S_{10}$ &$1+s+$  & $(1,1)$& $-2 $ &
    $0$ & $0$ & $0$ & $0$ &
    $-1$ & $-1$ & $(1,1)$ & $-{4\over 3}$ & $0$ \\
   $\overline S_{10}$ & $\alpha+2\gamma$  & $(1,1)$&$-2$ &
    $0$ & $0$ & $0$ & $0$ &
    $1$ & $1$ & $(1,1)$ & ${4\over 3}$ & $0$ \\
\hline
\end{tabular}
\label{sol2matter3}
\end{eqnarray*}
\caption{Massless states for solution II (\cite{top}).}
\end{table}
\end{document}